\documentclass[a4paper,12pt]{article}
\usepackage{epsfig}
\usepackage{axodraw}

\begin{document}


\newcommand{\bit}{\begin{itemize}}
\newcommand{\eit}{\end{itemize}}
\newcommand{\ti}{\textit}
\newcommand{\be}{\begin{equation}}
\newcommand{\ee}{\end{equation}}

\newcommand{\bc}{\begin{center}}
\newcommand{\ec}{\end{center}}
\newcommand{\nn}{\nonumber}
\newcommand{\f}{\frac}
\newcommand{\ba}{\begin{array}}
\newcommand{\ea}{\end{array}}
\def\stilde{\widetilde}
\def\gappeq{\mathrel{\rlap {\raise.5ex\hbox{$>$}}
{\lower.5ex\hbox{$\sim$}}}}
\def\lappeq{\mathrel{\rlap{\raise.5ex\hbox{$<$}}
{\lower.5ex\hbox{$\sim$}}}}
\newcommand{\rar}{\rightarrow}
\newcommand{\mb}{\mathbf}
\newcommand{\mc}{\mathcal}
\newcommand{\bo}{\mbox}
\newcommand{\veps}{\varepsilon}
\newcommand{\vthe}{\vtheta}
\newcommand{\proc}{$e^+e^-\rightarrow \tilde\chi_1^+\tilde\chi_1^-h^0$}
\newcommand{\accop}{$h^0 \tilde\chi_1^+\tilde\chi_1^-$}


\def\beq{\begin{equation}}
\def\eeq{\end{equation}}
\def\bea{\begin{eqnarray}}
\def\eea{\end{eqnarray}}
\newcommand{\ci}{{\cal I}}
\newcommand{\ca}{{\cal A}}
\newcommand{\Wp}{W^{\prime}}

\newcommand{\dedouble}{ \stackrel{ \leftrightarrow }{ \partial } }
\newcommand{\deR}{ \stackrel{ \rightarrow }{ \partial } }
\newcommand{\deL}{ \stackrel{ \leftarrow }{ \partial } }
\newcommand{\cp}{\tilde{\chi}^+_1}
\newcommand{\cm}{\tilde{\chi}^-_1}
\newcommand{\ccp}{\tilde{\chi}^+_2}
\newcommand{\ccm}{\tilde{\chi}^-_2}
\newcommand{\mcu}{m_{\tilde{\chi}^+_1}}
\newcommand{\mcd}{m_{\tilde{\chi}^+_2}}
\newcommand{\eepm}{e^+e^-}
\newcommand{\hcc}{e^+e^-\to h\; \tilde{\chi}^+_1\tilde{\chi}^-_1}
\newcommand{\hccc}{ h\, \tilde{\chi}^+_1\tilde{\chi}^-_1}
\newcommand{\hcccd}{ h\, \tilde{\chi}^+_1\tilde{\chi}^-_2}
\newcommand{\ahcu}{\alpha^2_{h\,\tilde{\chi}_1\tilde{\chi}_1}}
\newcommand{\ahcd}{\alpha^2_{h\,\tilde{\chi}_1\tilde{\chi}_2}}
\newcommand{\sqs}{\sqrt{s}}
\newcommand{\tb}{\tan{\beta}}
\newcommand{\msne}{M_{\tilde{\nu_e}}}
\newcommand{\sne}{\tilde{\nu_e}}
\newcommand{\maa}{m_{A^0}}
\newcommand{\mh}{m_h}
\newcommand{\dsdh}{E_h\frac{d\sigma}{d^3\mathbf{h}}}

\renewcommand{\thefootnote}{\fnsymbol{footnote}}
\rightline{ROME1-1378/2004}
\rightline{LC-TH-2004-012}
\vspace{.5cm} 
{\Large
\begin{center}
{\bf Associated production of a light Higgs boson \\ and a chargino pair  
in the MSSM \\ at linear colliders }

\end{center}}
\vspace{.3cm}

\begin{center}
Giancarlo Ferrera$^{1,2}\;$  and Barbara Mele$^{2,1}$ \\
\vspace{.3cm}
$^1$\emph{Dip. di Fisica, Universit\`a La Sapienza,
P.le A. Moro 2, I-00185 Rome, Italy}
\\
$^2$\emph{Istituto Nazionale di Fisica Nucleare, Sezione di Roma,
Rome, Italy\\
E-mail addresses :  {\small giancarlo.ferrera@roma1.infn.it,
barbara.mele@roma1.infn.it} 
}
\end{center}

\vspace{.3cm}
\hrule \vskip 0.3cm
\begin{center}
\small{\bf Abstract}\\[3mm]
\begin{minipage}[h]{14.0cm}
In the Minimal Supersymmetric Standard Model (MSSM), we study
the light Higgs-boson radiation off a light-chargino pair in
the process 
$\hcc$ at linear colliders with $\sqs=500$ GeV. 
We analyze cross sections in the regions of the
MSSM parameter space where the process $\hcc$ can not proceed
via  on-shell production and subsequent decay of either heavier charginos
or the pseudoscalar Higgs boson $A$. 
Cross sections up to a few
fb's are allowed, according to present experimental limits
on the Higgs-boson, chargino and sneutrino masses.
We also show how a measurement of the $\hcc$ production rate could provide a  
determination of the Higgs-boson couplings to charginos. 
\end{minipage}
\end{center}
\vskip 0.3cm \hrule \vskip 0.5cm
\section{Introduction}
Linear colliders would be a fantastic {\it precision instrument}
for Higgs boson physics and physics beyond the standard model (SM) 
that could show up at the LHC. In particular, if  supersymmetry (SUSY) 
exists with partners
of known particles  with masses not too far from present experimental
limits, a next-generation linear collider 
such as the International Linear Collider (ILC) 
\cite{Accomando:1997wt}
would be able to measure (sometimes with excellent precision) 
a number of crucial
parameters (such as masses, couplings and mixing angles),
and eventually test the fine structure of a particular SUSY model.
For instance, a linear collider at $\sqs=350$-500 GeV will be able 
to disentangle the characteristic two-doublet nature of a light
Higgs boson \cite{Gunion:1989we} of the Minimal Supersymmetric 
Standard Model (MSSM)  \cite{Nilles:1983ge,haber,barbieri}
even in the {\it decoupling} limit, where 
the light Higgs mimics the SM Higgs behavior, and all the other Higgs bosons
and SUSY partners are out of reach of both the LHC and linear colliders. 

Quite a few studies have been carried out to establish the linear-collider
potential in determining Higgs boson couplings to fermions, vector bosons, 
and also to SUSY partners \cite{Accomando:1997wt}. 
For coupling {\it suppressed} by the relatively light mass of the 
coupled particle
(as for the light fermions couplings to the Higgs bosons, 
where $g_{hf\bar f}\sim  m_f/v$),
the coupling is generally determined through the corresponding 
Higgs decay branching ratio measurement.

On the other hand, since the main Higgs production mechanisms
occur through the {\it unsuppressed} 
 Higgs boson couplings to vector bosons, 
the analysis of the Higgs boson
production cross sections is expected to provide a good determination of
the Higgs-bosons couplings to the $Z$ and $W$ vector bosons. 

Then, there are a number of couplings of the Higgs bosons
to quite heavy particles, other than gauge bosons,
that can not be investigated through Higgs boson decay channels
due to phase-space restrictions.
In the latter case, 
the associated production of a Higgs boson and a pair of
the heavy particles, when allowed by phase space, can provide
an alternative to measure  the
corresponding coupling. Some reduction  in the rate due to the 
possible phase-space saturation
by the heaviness of the final states is  expected in this case.

For instance, the SM Higgs-boson {\it unsuppressed} coupling to the top quark,
 $ m_t/v$,
can be determined at linear colliders with $\sqs\sim 1$TeV through
the production rates for the Higgs radiated off a top-quark pair
in the channel $e^+e^-\to h \; t \bar t$
\cite{Gaemers:1978jr}.

The latter strategy  can
be  useful  also in the MSSM, that introduces  an entire spectrum of 
relatively heavy partners, that in many cases are 
coupled to Higgs bosons via an {\it unsuppressed} coupling constant.

A typical example is that of the light Higgs-boson coupling to the
light top squark $h \;\tilde t_1 \tilde t_1$, that can be naturally large.
The continuum production $e^+e^-\to h \;\tilde t_1 \tilde t_1$ has been
studied in \cite{Belanger:1998rq} as a means of determining this coupling
(the corresponding channel at hadron colliders has been investigated
also in \cite{Djouadi:1997xx}).
Higgs-boson  production in association of sleptons and light neutralinos
in $e^+e^-$ collisions has been considered in \cite{Datta:2001sh}.

Following a similar strategy,
in the present work we want to investigate the possibility to measure
the light Higgs coupling to light charginos $\hccc$ through  the Higgs
boson production in association with a light-chargino pair
at linear colliders  
\beq
\hcc \; .
\label{duno}
\eeq
 Note  that
 {\it heavy} Higgs bosons couplings to SUSY partners can be mostly explored
 via  Higgs decay rates. 
For instance, heavy Higgs decays into chargino/neutralino pairs 
 and sfermion pairs in the MSSM 
have been reviewed  in \cite{Djouadi:1996pj}.
The {\it precision} measurement of the Higgs-chargino couplings
at a muon collider operating at a heavy Higgs boson resonance
has been discussed in \cite{Fraas:2003cx}.
On the other hand,
as far as the {\it light} Higgs boson coupling to light charginos is concerned, 
not much can be learned 
through Higgs decay channels due to  phase-space restrictions.
Indeed, in the MSSM $m_h$ is expected to be lighter that about 
130 GeV \cite{Brignole:2001jy},
and the present experimental limit on the chargino mass
 $m_{\cp}>103.5$ GeV 
 (or even the milder one $m_{\cp}>92.4$ GeV, in case of almost degenerate
 chargino and lightest neutralino)
 \cite{LEPSUSYWG} excludes the decay $h\to\cp\cm$.

Hence, the simplest way to determine the $\hccc$ coupling could be through
the measurement of the rate for the light Higgs-boson production 
at linear colliders in the channel
$\hcc$. The present mass 
limits allow a good potential 
for covering a considerable area of the MSSM parameter space,
even at  $\sqs\simeq 500$ GeV.

We will  concentrate on the {\it non resonant} continuum production
$\hcc\;$,
that is, we will not include in our study the cases where the 
considered process
proceeds through the on-shell production of 
either a $\cp \tilde{\chi}^-_2$ (or the charged conjugated $\cm
\tilde{\chi}^+_2$) or  a $hA$ 
intermediate state
(where $\tilde{\chi}^-_2$
is the heavier chargino and $A$ is the
pseudoscalar Higgs boson) with a subsequent decay
$\tilde{\chi}^-_2 \to h \cm$ and $A\to\cp\cm$, respectively.
In the latter cases, the total $\hccc$
production rates are in general enhanced with respect to the 
continuum
production, that can be viewed as a higher-order  process in the 
electroweak coupling.
We will also assume  either a low value 
(i.e., $\msne~100$ GeV) or a quite large value (i.e., $\msne~500$ GeV)
for the electron sneutrino mass. The latter suppresses the 
Feynman diagrams with a
sneutrino exchange, involving predominantly the gaugino
components of the light charginos.

Note that the SM process $\eepm \to H W^+W^-$  
(that can be somehow connected by a SuSy
transformation to $\hcc$)  has a total 
cross section of about 5.6 fb  for $m_H\simeq 120$ GeV,
at $\sqs\simeq 500$ GeV \cite{Baillargeon:1993iw}.

The measurement of the $\hccc$ coupling through the process
$\hcc$ would complement the nice set of precision  measurements
in the chargino sector expected at future high energy colliders 
(see \cite{lesh} and reference therein).

The plan of the paper is the following. In Section 2, the MSSM parameter
regions that are of relevance for the non resonant $\hccc$ production
are discussed.
We also define three {\it reference scenarios} for the following analysis.
In Section 3, the matrix element for $\hcc$ is presented,
and the cross-section computation is described.
In Section 4, we present total cross sections versus the MSSM parameters.
In Section 5, we discuss the foreseen sensitivity to a determination
of the $\hccc$ coupling on a event-number basis,
before giving our conclusions in Section 6.
In Appendix A, we define the interaction Lagrangian and couplings.
In Appendix B, we describe the phase-space integration of the relevant
squared  matrix elements. 
\unitlength=1.0 pt
\SetScale{1.0}
\SetWidth{0.7}      
%
\begin{figure}[ht!]
\begin{center}
\begin{picture}(160,130)(0,0)
\Text(13.0,90.0)[r]{$e^+$}
\ArrowLine(55,60)(15,90) 
\Text(40.0,82.0)[]{$p_1$}
\Text(13.0,30.0)[r]{$e^-$}
\ArrowLine(15,30)(55,60) 
\Text(40.0,38.0)[]{$p_2$}
\Text(72.0,55.0)[t]{$\gamma$}
\Photon(55.0,60.0)(100.0,60.0){3}{4} 
\Text(72.0,69.0)[]{$k$}
\Text(142.0,30.0)[l]{$\tilde\chi^-_1$}
\ArrowLine(140.0,30.0)(100.0,60.0) 
\Text(120.0,38.0)[]{$q_2$}
\Text(98.0,75.0)[r]{$\tilde\chi^+_1$}
\ArrowLine(100.0,60.0)(100.0,90.0) 
\Text(108.0,75.0)[]{$q_3$}
\Text(142.0,110)[l]{$\tilde\chi^+_1$}
\ArrowLine(100.0,90.0)(140.0,110)
\Text(120.0,107.0)[]{$q_1$}

\Text(142.0,70.0)[l]{$h^0$}
\DashLine(100.0,90.0)(140.0,70.0){2.5} 
\Text(120.0,87.0)[]{$h$}
\Text(77.5,0)[b] {$A_1$}
\end{picture} \ 
%
\hspace{2cm}
%
\begin{picture}(160,130)(0,0)
\Text(13.0,90.0)[r]{$e^+$}
\ArrowLine(55,60)(15,90) 
\Text(40.0,82.0)[]{$p_1$}
\Text(13.0,30.0)[r]{$e^-$}
\ArrowLine(15,30)(55,60) 
\Text(40.0,38.0)[]{$p_2$}
\Text(72.0,55.0)[t]{$\gamma$}
\Photon(55.0,60.0)(100.0,60.0){3}{4} 
\Text(72.0,69.0)[]{$k$}
\Text(141.0,90.0)[l]{$\tilde\chi^+_1$}
\ArrowLine(100,60)(140,90) 
\Text(120.0,83.0)[]{$q_2$}
\Text(98.0,45.0)[r]{$\tilde\chi^-_1$}
\ArrowLine(100.0,30.0)(100.0,60.0) 
\Text(108.0,45.0)[]{$q_4$}
\Text(142.0,50.0)[l]{$h^0$}
\DashLine(100.0,30.0)(140.0,50.0){2.5} 
\Text(120.0,34.0)[]{$h$}
\Text(142.0,10)[l]{$\tilde\chi^-_1$}
\ArrowLine(140,10)(100,30)
\Text(120.0,13.0)[]{$q_1$}
\Text(77.5,0)[b] {$A_2$}
\end{picture} \ 
\newline
\begin{picture}(160,130)(0,0)
\Text(13.0,90.0)[r]{$e^+$}
\ArrowLine(55,60)(15,90) 
\Text(40.0,82.0)[]{$p_1$}
\Text(13.0,30.0)[r]{$e^-$}
\ArrowLine(15,30)(55,60) 
\Text(40.0,38.0)[]{$p_2$}
\Text(72.0,55.0)[t]{$Z^0$}
\Photon(55.0,60.0)(100.0,60.0){3}{4} 
\Text(72.0,69.0)[]{$k$}
\Text(142.0,30.0)[l]{$\tilde\chi^-_1$}
\ArrowLine(140.0,30.0)(100.0,60.0) 
\Text(120.0,38.0)[]{$q_2$}
\Text(98.0,75.0)[r]{$\tilde\chi^+_1$}
\ArrowLine(100.0,60.0)(100.0,90.0) 
\Text(108.0,75.0)[]{$q_3$}
\Text(142.0,110)[l]{$\tilde\chi^+_1$}
\ArrowLine(100.0,90.0)(140.0,110)
\Text(120.0,107.0)[]{$q_1$}
\Text(142.0,70.0)[l]{$h^0$}
\DashLine(100.0,90.0)(140.0,70.0){2.5} 
\Text(120.0,87.0)[]{$h$}
\Text(77.5,0)[b] {$A_3$}
\end{picture} \ 
%
\hspace{2cm}
\begin{picture}(160,130)(0,0)
\Text(13.0,90.0)[r]{$e^+$}
\ArrowLine(55,60)(15,90) 
\Text(40.0,82.0)[]{$p_1$}
\Text(13.0,30.0)[r]{$e^-$}
\ArrowLine(15,30)(55,60) 
\Text(40.0,38.0)[]{$p_2$}
\Text(72.0,55.0)[t]{$Z^0$}
\Photon(55.0,60.0)(100.0,60.0){3}{4} 
\Text(72.0,69.0)[]{$k$}
\Text(141.0,90.0)[l]{$\tilde\chi^+_1$}
\ArrowLine(100,60)(140,90) 
\Text(120.0,83.0)[]{$q_2$}
\Text(98.0,45.0)[r]{$\tilde\chi^-_1$}
\ArrowLine(100.0,30.0)(100.0,60.0) 
\Text(108.0,45.0)[]{$q_4$}
\Text(142.0,50.0)[l]{$h^0$}
\DashLine(100.0,30.0)(140.0,50.0){2.5} 
\Text(120.0,34.0)[]{$h$}
\Text(142.0,10)[l]{$\tilde\chi^-_1$}
\ArrowLine(140,10)(100,30)
\Text(120.0,13.0)[]{$q_1$}
\Text(77.5,0)[b] {$A_4$}
\end{picture} \ 
\newline
\begin{picture}(160,130)(0,0)
\Text(13.0,90.0)[r]{$e^+$}
\ArrowLine(55,60)(15,90) 
\Text(40.0,82.0)[]{$p_1$}
\Text(13.0,30.0)[r]{$e^-$}
\ArrowLine(15,30)(55,60) 
\Text(40.0,38.0)[]{$p_2$}
\Text(72.0,55.0)[t]{$Z^0$}
\Photon(55.0,60.0)(100.0,60.0){3}{4} 
\Text(72.0,69.0)[]{$k$}
\Text(142.0,30.0)[l]{$\tilde\chi^-_1$}
\ArrowLine(140.0,30.0)(100.0,60.0) 
\Text(120.0,38.0)[]{$q_2$}
\Text(98.0,75.0)[r]{$\tilde\chi^+_2$}
\ArrowLine(100.0,60.0)(100.0,90.0) 
\Text(108.0,75.0)[]{$q_3$}
\Text(142.0,110)[l]{$\tilde\chi^+_1$}
\ArrowLine(100.0,90.0)(140.0,110)
\Text(120.0,107.0)[]{$q_1$}
\Text(142.0,70.0)[l]{$h^0$}
\DashLine(100.0,90.0)(140.0,70.0){2.5} 
\Text(120.0,87.0)[]{$h$}
\Text(77.5,0)[b] {$A_5$}
\end{picture} \ 
%
\hspace{2cm}
\begin{picture}(160,130)(0,0)
\Text(13.0,90.0)[r]{$e^+$}
\ArrowLine(55,60)(15,90) 
\Text(40.0,82.0)[]{$p_1$}
\Text(13.0,30.0)[r]{$e^-$}
\ArrowLine(15,30)(55,60) 
\Text(40.0,38.0)[]{$p_2$}
\Text(72.0,55.0)[t]{$Z^0$}
\Photon(55.0,60.0)(100.0,60.0){3}{4} 
\Text(72.0,69.0)[]{$k$}
\Text(141.0,90.0)[l]{$\tilde\chi^+_1$}
\ArrowLine(100,60)(140,90) 
\Text(120.0,83.0)[]{$q_2$}
\Text(98.0,45.0)[r]{$\tilde\chi^-_2$}
\ArrowLine(100.0,30.0)(100.0,60.0) 
\Text(108.0,45.0)[]{$q_4$}
\Text(142.0,50.0)[l]{$h^0$}
\DashLine(100.0,30.0)(140.0,50.0){2.5} 
\Text(120.0,34.0)[]{$h$}
\Text(142.0,10)[l]{$\tilde\chi^-_1$}
\ArrowLine(140,10)(100,30)
\Text(120.0,13.0)[]{$q_1$}
\Text(77.5,0)[b] {$A_6$}
\end{picture} \ 
%
\newline
\hspace{2cm}
\begin{picture}(160,130)(0,0)
\Text(13.0,90.0)[r]{$e^+$}
\ArrowLine(55,60)(15,90) 
\Text(40.0,82.0)[]{$p_1$}
\Text(13.0,30.0)[r]{$e^-$}
\ArrowLine(15,30)(55,60) 
\Text(40.0,38.0)[]{$p_2$}
\Text(72.0,55.0)[t]{$Z^0$}
\Photon(55.0,60.0)(100.0,60.0){3}{4} 
\Text(72.0,69.0)[]{$k$}
\Text(141.0,90.0)[l]{$h^0$}
\DashLine(100,60)(140,90){2.5} 
\Text(120.0,83.0)[]{$h$}
\Text(98.0,45.0)[r]{$Z^0$}
\Photon(100.0,30.0)(100.0,60.0){3}{3} 
\Text(107.0,45.0)[]{$q$}
\Text(142.0,50.0)[l]{$\tilde\chi+_1$}
\ArrowLine(100.0,30.0)(140.0,50.0) 
\Text(120.0,33.0)[]{$q_2$}
\Text(142.0,10)[l]{$\tilde\chi^-_1$}
\ArrowLine(140,10)(100,30)
\Text(120.0,13.0)[]{$q_1$}
\Text(77.5,0)[b] {$A_7$}
\end{picture} \ 
\hspace{2cm}
\begin{picture}(160,130)(0,0)
\Text(13.0,90.0)[r]{$e^+$}
\ArrowLine(55,60)(15,90) 
\Text(40.0,82.0)[]{$p_1$}
\Text(13.0,30.0)[r]{$e^-$}
\ArrowLine(15,30)(55,60) 
\Text(40.0,38.0)[]{$p_2$}
\Text(72.0,55.0)[t]{$Z^0$}
\Photon(55.0,60.0)(100.0,60.0){3}{4} 
\Text(72.0,69.0)[]{$k$}
\Text(141.0,90.0)[l]{$h^0$}
\DashLine(100,60)(140,90){2.5} 
\Text(120.0,83.0)[]{$h$}
\Text(98.0,45.0)[r]{$A^0$}
\DashLine(100.0,30.0)(100.0,60.0){2.5} 
\Text(107.0,45.0)[]{$q$}
\Text(142.0,50.0)[l]{$\tilde\chi^+_1$}
\ArrowLine(100.0,30.0)(140.0,50.0) 
\Text(120.0,33.0)[]{$q_2$}
\Text(142.0,10)[l]{$\tilde\chi^-_1$}
\ArrowLine(140,10)(100,30)
\Text(120.0,13.0)[]{$q_1$}
\Text(77.5,0)[b] {$A_8$}
\end{picture}
\caption{{\small
Set of $s-$channel Feynman diagrams contributing to $\hcc$.
 }}
\label{uno}
\end{center}
\end{figure}
\begin{figure}[ht!]
\begin{center}
{} \qquad\allowbreak
\hspace{2cm}
\begin{picture}(160,130)(0,0)
\Text(13.0,90.0)[r]{$e^+$}
\ArrowLine(55,90)(15,90) 
\Text(40.0,97.0)[]{$p_1$}
\Text(13.0,30.0)[r]{$e^-$}
\ArrowLine(15,30)(55,30) 
\Text(40.0,22.0)[]{$p_2$}
\Text(51,60)[r]{$\tilde\nu$}
\DashArrowLine(55,30)(55,90){2.5}
\Text(65.0,60.0)[]{$q_5$}
\Text(142,90)[l]{$\tilde\chi^+_1$}
\ArrowLine(55,90)(140,90)
\Text(80.0,22.0)[]{$q_4$}
\Text(77.5,33)[b]{$\tilde\chi^+_1$}
\ArrowLine(100,30)(55,30)
\Text(97.0,97.0)[]{$q_2$}
\Text(142.0,50.0)[l]{$h^0$}
\DashLine(100.0,30.0)(140.0,50.0){2.5} 
\Text(120.0,33.0)[]{$h$}
\Text(142.0,10)[l]{$\tilde\chi^-_1$}
\ArrowLine(140,10)(100,30)
\Text(120.0,13.0)[]{$q_1$}
\Text(77.5,0)[b] {$A_9$}
\end{picture} \ 
\newline
\begin{picture}(160,130)(0,0)
\Text(13.0,90.0)[r]{$e^+$}
\ArrowLine(55,90)(15,90) 
\Text(40.0,97.0)[]{$p_1$}
\Text(13.0,30.0)[r]{$e^-$}
\ArrowLine(15,30)(55,30) 
\Text(40.0,22.0)[]{$p_2$}
\Text(51,60)[r]{$\tilde\nu$}
\DashArrowLine(55,30)(55,90){2.5}
\Text(65.0,60.0)[]{$q_5$}
\Text(142,90)[l]{$\tilde\chi^+_1$}
\ArrowLine(55,90)(140,90)
\Text(80.0,22.0)[]{$q_4$}
\Text(77.5,33)[b]{$\tilde\chi^+_2$}
\ArrowLine(100,30)(55,30)
\Text(97.0,97.0)[]{$q_2$}
\Text(142.0,50.0)[l]{$h^0$}
\DashLine(100.0,30.0)(140.0,50.0){2.5} 
\Text(142.0,10)[l]{$\tilde\chi^-_1$}
\ArrowLine(140,10)(100,30)
\Text(120.0,13.0)[]{$q_1$}
\Text(77.5,0)[b] {$A_{10}$}
\end{picture} \
{} \qquad\allowbreak
\hspace{2cm}
\begin{picture}(160,130)(0,0)
\Text(13.0,90.0)[r]{$e^+$}
\ArrowLine(55,90)(15,90) 
\Text(40.0,97.0)[]{$p_1$}
\Text(13.0,30.0)[r]{$e^-$}
\ArrowLine(15,30)(55,30) 
\Text(40.0,22.0)[]{$p_2$}
\Text(51,60)[r]{$\tilde\nu$}
\DashArrowLine(55,30)(55,90){2.5}
\Text(65.0,60.0)[]{$q_6$}
\Text(142,30)[l]{$\tilde\chi^-_1$}
\ArrowLine(140,30)(55,30)
\Text(97.0,22.0)[]{$q_1$}
\Text(77.5,77)[b]{$\tilde\chi^+_1$}
\ArrowLine(55,90)(100,90)
\Text(80.0,97.0)[]{$q_3$}
\Text(142.0,110.0)[l]{$h^0$}
\DashLine(100.0,90.0)(140.0,110.0){2.5} 
\Text(120.0,107.0)[]{$h$}
\Text(142.0,70)[l]{$\tilde\chi^+_1$}
\ArrowLine(100,90)(140,70)
\Text(120.0,87.0)[]{$q_2$}
\Text(77.5,0)[b] {$A_{11}$}
\end{picture} \ 
\newline 
\begin{picture}(160,130)(0,0)
\Text(13.0,90.0)[r]{$e^+$}
\ArrowLine(55,90)(15,90) 
\Text(40.0,97.0)[]{$p_1$}
\Text(13.0,30.0)[r]{$e^-$}
\ArrowLine(15,30)(55,30) 
\Text(40.0,22.0)[]{$p_2$}
\Text(51,60)[r]{$\tilde\nu$}
\DashArrowLine(55,30)(55,90){2.5}
\Text(65.0,60.0)[]{$q_6$}
\Text(142,30)[l]{$\tilde\chi^-_1$}
\ArrowLine(140,30)(55,30)
\Text(97.0,22.0)[]{$q_1$}
\Text(77.5,77)[b]{$\tilde\chi^+_2$}
\ArrowLine(55,90)(100,90)
\Text(80.0,97.0)[]{$q_3$}
\Text(142.0,110.0)[l]{$h^0$}
\DashLine(100.0,90.0)(140.0,110.0){2.5} 
\Text(120.0,107.0)[]{$h$}
\Text(142.0,70)[l]{$\tilde\chi^+_1$}
\ArrowLine(100,90)(140,70)
\Text(120.0,87.0)[]{$q_2$}
\Text(77.5,0)[b] {$A_{12}$}
\end{picture} \ 
{} \qquad\allowbreak
\hspace{2cm}
\begin{picture}(160,130)(0,0)
\Text(13.0,90.0)[r]{$e^+$}
\ArrowLine(77.5,90)(15,90) 
\Text(48.0,97.0)[]{$p_1$}
\Text(13.0,30.0)[r]{$e^-$}
\ArrowLine(15,30)(77.5,30) 
\Text(48.0,22.0)[]{$p_2$}
\Text(74,75)[r]{$\tilde\nu$}
\DashArrowLine(77.5,60)(77.5,90){2.5}
\Text(87.0,75.0)[]{$q_5$}
\Text(74,45)[r]{$\tilde\nu$}
\DashArrowLine(77.5,30)(77.5,60){2.5}
\Text(87.0,45.0)[]{$q_6$}
\Text(142,90)[l]{$\tilde\chi^+_1$}
\ArrowLine(77.5,90)(140,90)
\Text(112.0,97.0)[]{$q_2$}
\Text(142,30)[l]{$\tilde\chi^-_1$}
\ArrowLine(140,30)(77.5,30)
\Text(112.0,22.0)[]{$q_1$}
\Text(142.0,60)[l]{$h^0$}
\DashLine(77.5,60)(140.0,60){2.5} 
\Text(112.0,65.0)[]{$h$}
\Text(77.5,0)[b] {$A_{13}$}
\end{picture} \ 
{} \qquad\allowbreak
\caption{{\small 
Set of $t-$channel Feynman diagrams contributing to $\hcc$.
}}
\label{due}
\end{center}
\end{figure}
\section{Relevant MSSM Parameter Space}
Charginos are expected to be in general among the lightest SuSy partners
in the new particle spectrum of the MSSM. This makes interesting to consider
the production of a light Higgs boson associated to two light charginos
in the process $\hcc$ at $\sqs=500$ GeV,
even if all the particles in the final states are expected to be 
 not so light, and in general heavier than 
100 GeV. \\
Charginos are the mass eigenstates of the mass matrix 
that mixes charged gaugino and 
higgsino states (see \cite{haber}, and Appendix A). 
At tree level, the latter depends on three parameters,
$M_2$, $\mu$ and $\tb$. When the mass matrix is real, the two diagonalizing
matrices can be expressed in terms of two mixing angles, $\phi_{\pm}$.
Then, the mass eigenvalues $\mcu$ and $\mcd$ and the mixing angles can be 
analytically written in terms of the parameters $M_2$, $\mu$ and $\tb$. 
The presence of a Higgs boson in the process $\hcc$ requires
at tree level a further
parameter, that can be the pseudoscalar mass $\maa$. 
On the other hand, the inclusion of the main radiative corrections to 
the Higgs boson mass
\begin{figure}[th]
\vspace{-0.7cm}
\centerline{\epsfxsize=4.20truein \epsfbox{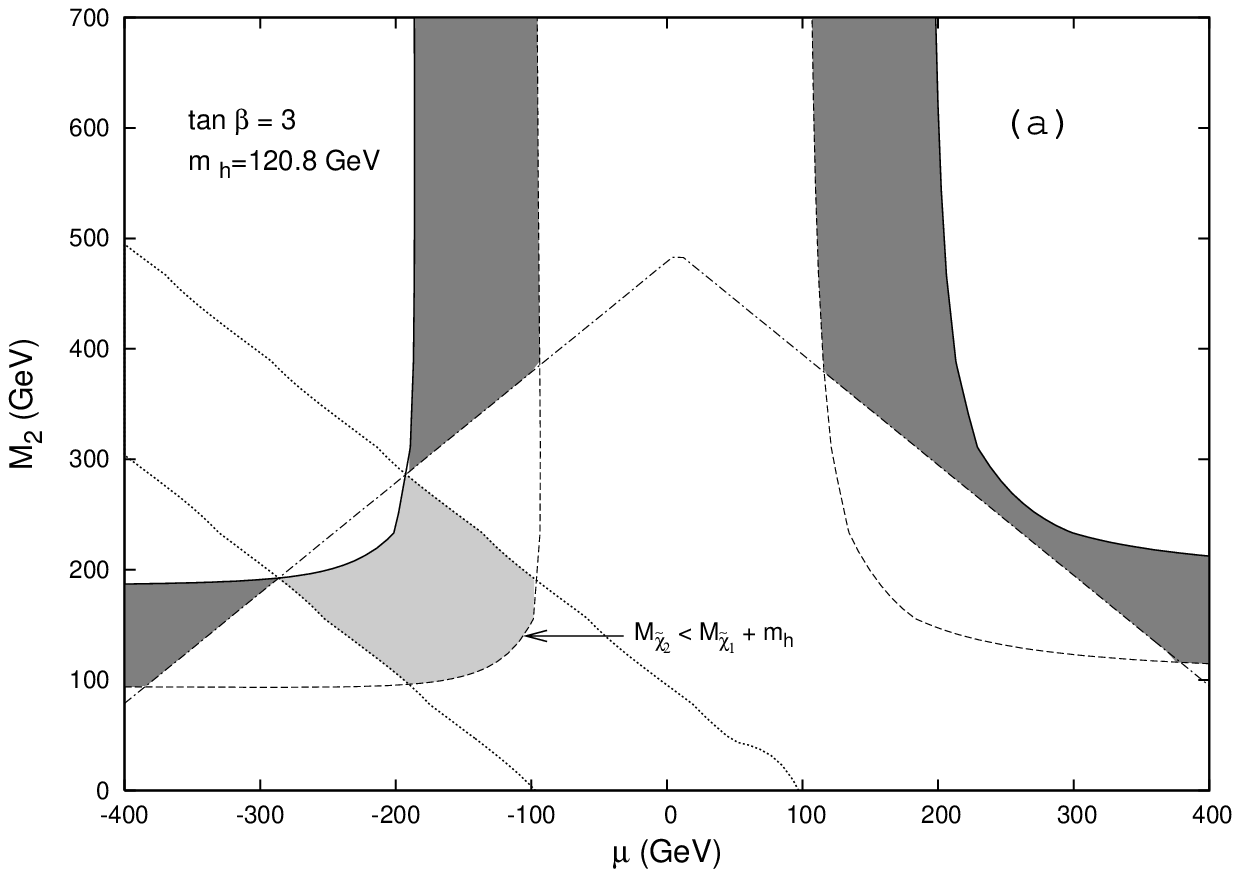}}
\centerline{\epsfxsize=4.20truein \epsfbox{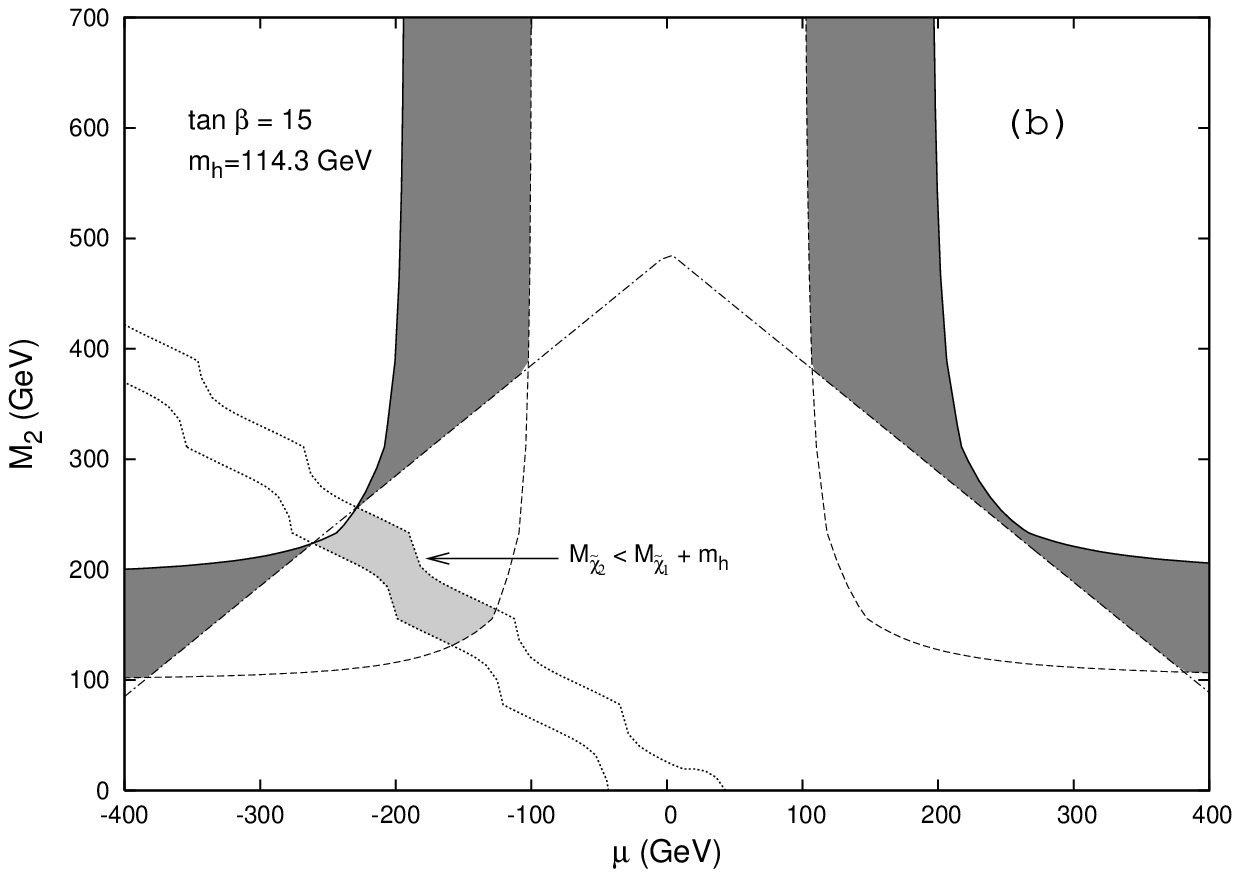}}
\centerline{\epsfxsize=4.20truein \epsfbox{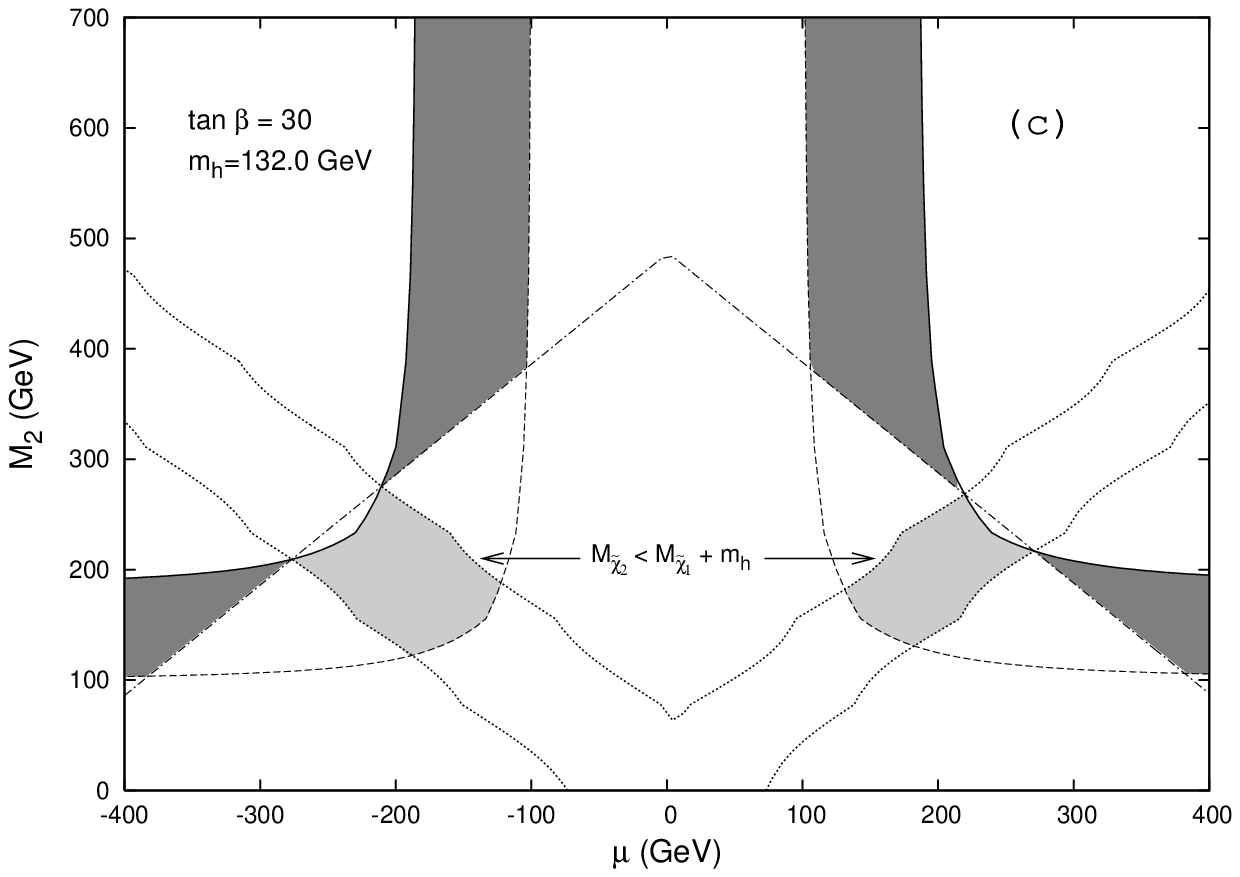}}
\caption{
\small 
MSSM parameter regions allowed for the continuum production
$\hcc$ at $\sqrt s=500~GeV$, for $\tan\beta=3,15,30$ and $\maa=500~GeV$
(in either light or dark grey). 
}
\label{tre}
\end{figure}
and couplings
involves all the basic parameters needed for 
setting the complete
mass spectrum  of the SuSy partners in the MSSM.
In our study of $\hcc$ at $\sqs=$500 GeV, we set $\maa=500$ GeV. This
pushes the pseudoscalar field $A^0$ beyond the threshold for direct production,
thus preventing resonant $A^0\to \cp\cm$ contribution to the $\hccc$ final
state.
At the same time, such a large value for $\maa$ 
sets a {\it decoupling-limit} scenario ($\maa\gg M_Z$). 

Present experimental lower limits
on $\mh$ \cite{lephiggsMSSM} in the {\it decoupling-limit} MSSM  
are close to the ones derived from  the SM Higgs boson direct search
(i.e., $m_H>114.4$ GeV at 95\% C.L. \cite{lephiggsSM}).

The corrections to the light Higgs mass and  coupling parameter 
$\alpha$\footnote{The inclusion of radiative corrections to the 
Higgs-boson coupling
would require in principle a more general treatment of the 
complete set of radiative corrections to the process under consideration.
On the other hand, one can see that the simple inclusion of the correction 
to the parameter $\alpha$ is to a good extent self consistent
in our case. The latter has anyway a minor impact
on our results.}
(cf. Appendix~A)
have been computed according to the code FeynHiggsFast \cite{FeynHiggsFast}, 
with the following  input
parameters :  
$M_{\tilde{t}_{L,R}}=M_{\tilde{b}_{L,R}}=M_{\tilde{g}}=1$ TeV , 
$X_t\;(\equiv A_t-\mu \cot{\beta})=$ either 0 or 2 TeV,
$A_b=A_t$, $m_t=175$ GeV, $m_b=4.5$ GeV, $\mu=200$ GeV, $M_2=400$ GeV, 
and renormalization scale at $m_t$, in the most complete version 
of the code 
\footnote{Varying the $\mu$ and $M_2$
parameters 
would affect the Higgs spectrum and couplings negligibly.}.

Then, in our study, we assumed  three different $\tb$ scenarios, 
and corresponding $\mh$ values for $\maa=500$ GeV:

\vspace{0.2cm}
\noindent
{\bf a)} $\tb=3$, with {\it maximal} stop mixing 
(i.e., $X_t=2$ TeV), 
and $\mh=120.8$ GeV;

\noindent
{\bf b)} $\tb=15$, with {\it no} stop mixing (i.e., $X_t=0$), 
and $\mh=114.3$ GeV;

\noindent
{\bf c)} $\tb=30$, with {\it maximal} stop mixing (i.e., $X_t=2$ TeV), 
and $\mh=132.0$ GeV; 

\vspace{0.2cm}
\noindent
that are allowed by present experimental limits \cite{lephiggsMSSM}.

\vspace{0.3cm}
\noindent
The 13 Feynman diagrams corresponding to the
process $\hcc$ arise either from 
the $s$-channel $Z^0$/$\gamma$ exchange (cf. Fig.~\ref{uno})
or from the $t$-channel 
electron-sneutrino $\sne$ exchange (cf. Fig.~\ref{due}).
Hence, $\msne$ is a further crucial parameter in the present analysis,
influencing the relative importance of $t$-channel diagrams.

In our cross-section evaluation, we include 
all the 13 diagrams.

In Fig.~\ref{tre}, we show (in either light or dark grey), 
the area in the $(\mu,M_2)$ plane that is of relevance for the 
{\it non resonant}
$\hcc$ process, for the three different $\tb$ scenarios.
The solid lines correspond to the threshold energy contour level :
\beq
\sqs=2\; \mcu +\mh ,
\label{sogliauno}
\eeq 
while the dashed lines  refer to the experimental limit on the 
light chargino mass ($\mcu\simeq 100$ GeV).

The straight dot-dashed lines limit from above the region that allows
the associated production of a light chargino $\cp$ and
a resonant heavier chargino $\ccm$
 (that we are not interested in), and correspond to :
\beq
\sqs=\mcu +m_{\ccm} .
\label{sogliadue}
\eeq 
 
A further region of interest (beyond the dark-grey one) is the one
where, although $\sqs>\mcu +m_{\ccm}$, the heavier chargino is {\it below}
the threshold for a direct decay $\ccp \to \cp h$.
Then, again, a resonant $\ccp$  is not  allowed.
The area where $\mcd<\mcu+\mh$ is the one inside the oblique
stripes in Fig.~\ref{tre}.
The intersection of these stripes with the area between the
solid and dashed curves (light-grey regions) 
is a further  region
relevant to the non resonant $\hcc$ process. 

We stress that the constraints on the MSSM parameter space shown in 
Fig.~\ref{tre} are purely of {\it kinematical} nature.

On the other hand, the dynamical (coupling) characteristics of our 
process will also derive from the MSSM parameters. For example,
it is well known that, in regions where $|\mu|\gg M_2$, the gaugino component
in the light  charginos is dominant (enhancing the coupling to the sneutrino
in the $t$-channel diagrams
in Fig.~\ref{due}), while for $M_2 \gg |\mu|$ light charginos behave mostly
like higgsinos (enhancing the couplings to $Z/\gamma$ in the $s$-channel 
diagrams in Fig.~\ref{uno}). 

\begin{figure}[th]
\vspace{-1.5cm}
\centerline{\epsfxsize=6.truein \epsfbox{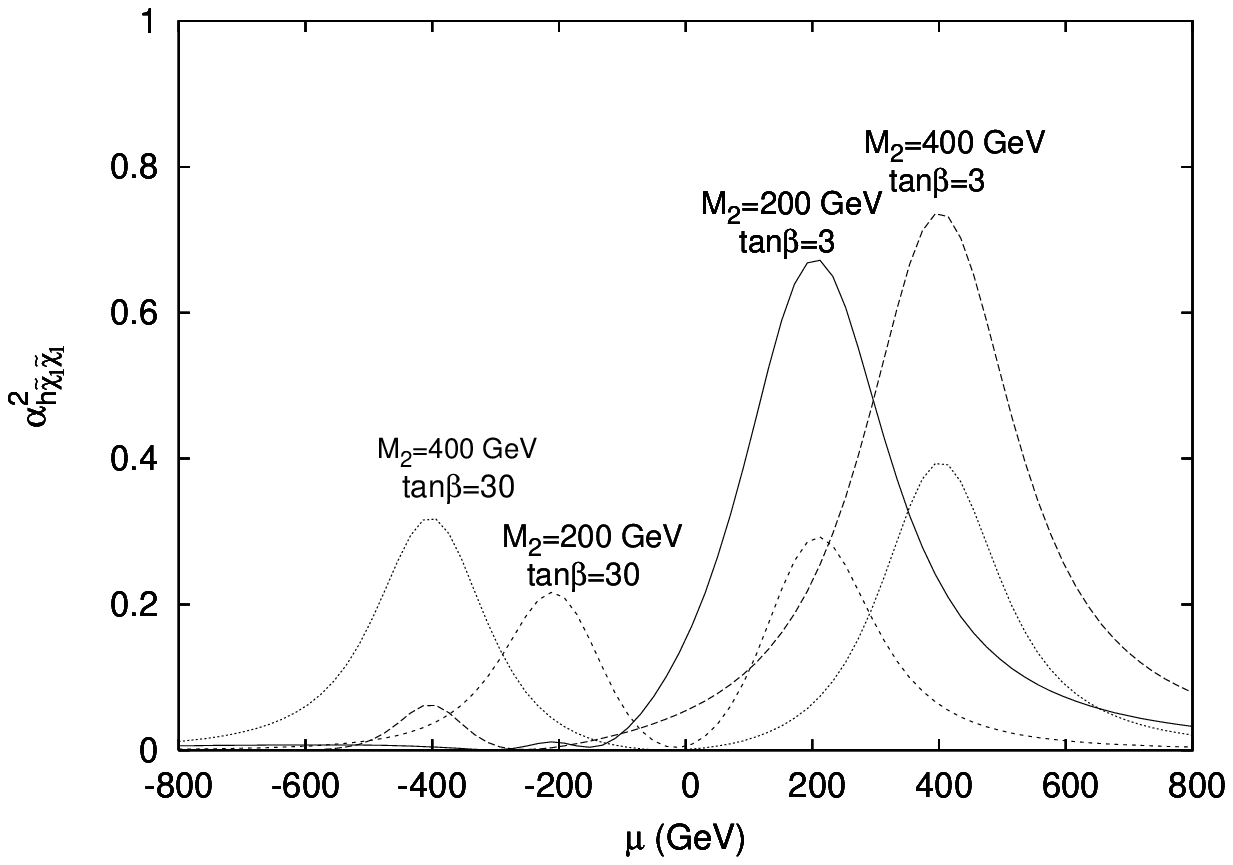}}
\vspace{0.3cm}
\centerline{\epsfxsize=6.truein \epsfbox{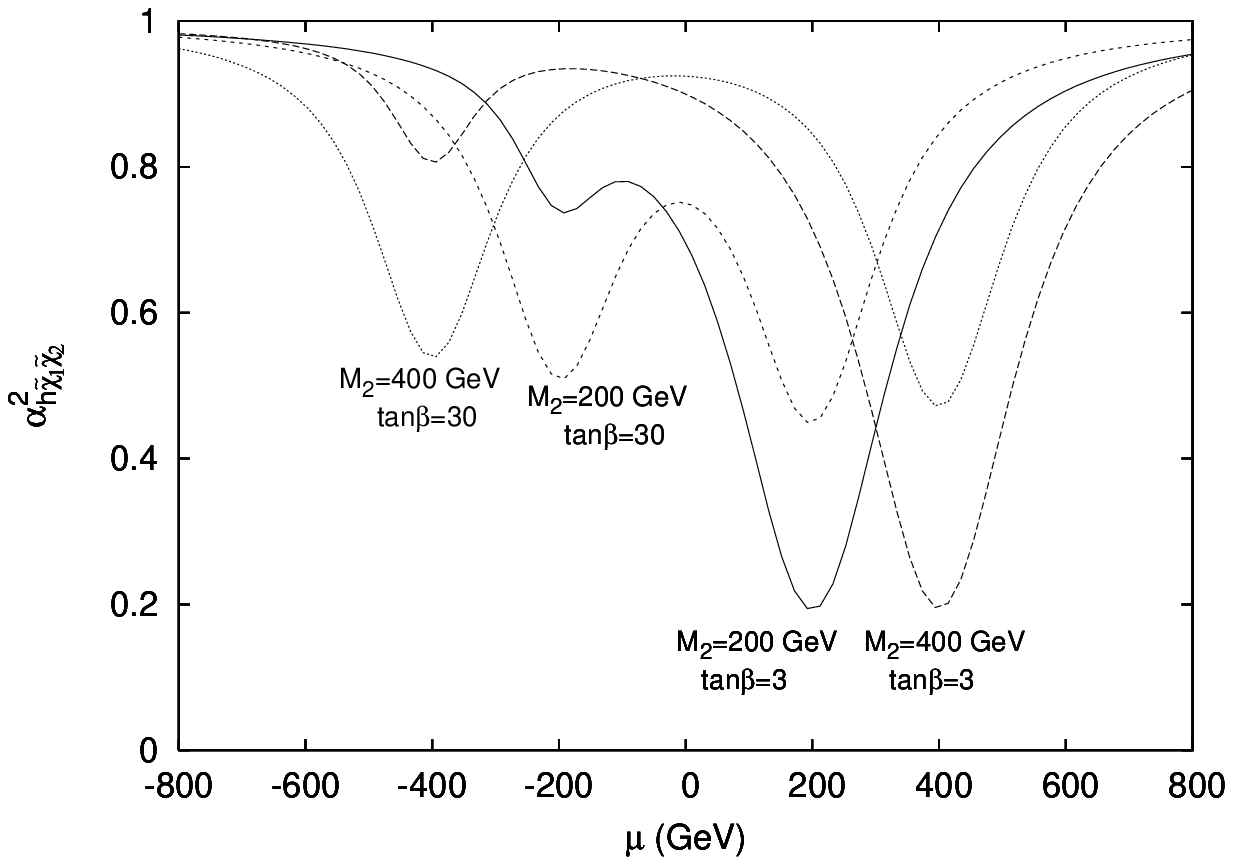}}
\caption{
\small
Squared couplings for $\hccc$ (upper plot) and $\hcccd$ (lower plot),
as defined by Eqs.~(\ref{coup-uno}) and (\ref{coup-due}) in the text, 
respectively .
}
\label{couplings}
\end{figure}

Since we are particularly interested  to a possible determination of the
$\hccc$ coupling,
in Fig.~\ref{couplings} (upper part) we show the behavior of the squared
$\hccc$  coupling, versus $\mu$, at fixed $M_2$ and $\tb$.
In particular, we define
\beq
\ahcu \equiv |C_{11}^{L}|^2 + |C_{11}^{R}|^2 = 2|C_{11}^{L}|^2\;,
\label{coup-uno}
\eeq   
where $C_{ij}^{L,R}$ are defined in Appendix A, 
by Eqs.~(\ref{cdefil}) and (\ref{cdefir}).
\\
Fig.~\ref{couplings}
shows clearly that the $\hccc$ coupling is
maximized for $\mu\simeq M_2$. A second local maximum, 
that is more pronounced at large $\tb$ values, occurs 
at $\mu\simeq - M_2$. On the other hand,
a ratio $M_2/|\mu|$ quite different from 1 
(corresponding to the dominance of either the
gaugino or the higgsino component in the $\cp$)
implies in general a  depleted
$\hccc$ coupling.
\clearpage
One can then confront the $\hccc$ coupling enhancement condition
$|\mu|\simeq M_2$ with 
the allowed parameter space for $\hcc$ in Fig.~\ref{tre}.
The light-grey region (corresponding to $\mcd<\mcu+\mh$) is
characterized by a local enhancement of the $\hccc$ coupling,
that is more pronounced at positive $\mu$ (only allowed at large $\tb$).
Instead, in most of the dark-grey region, one has a moderate
value of the $\hccc$ coupling.

On the other hand,
one can note that the parameter dependence of the  $\hcccd$ coupling
 (entering the amplitudes
$A_5, A_6$ in Fig.~\ref{uno} and $A_{10}, A_{12}$ in Fig.~\ref{due}),
that involves the heavier chargino,
is almost complementary to the
$\hccc$ one. This is clearly shown in the lower part of
Fig.~\ref{couplings}, where we define
\beq
\ahcd \equiv |C_{12}^{L}|^2 + |C_{12}^{R}|^2 = \\
|C_{21}^{R}|^2 + |C_{21}^{L}|^2 \; .
\label{coup-due}
\eeq  
Indeed, the $\hcccd$ coupling tends to be maximal for most of the parameter 
values, apart from the regions where  $M_2/|\mu|\sim 1$.

The fact that a large $\hccc$ coupling implies $M_2/|\mu|\sim 1$ (that  is
  substantial
components of both gaugino and higgsino in the lightest chargino) makes both
$s-$ and $t-$channel amplitudes relevant for the  coupling analysis.
This, joined to the complementarity of the $\hccc$ and $\hcccd$ couplings,
makes the behavior of the production cross sections for $\hcc$ in terms of the
fundamental MSSM parameters not always easy to interpret.
\\
For this reason, here we  study the $\hccc$ production rate
through a  choice of basic parameters differing from the usual one,
and affecting  the cross-section behavior in a more transparent way. 
Apart from
$\tb$ and the sneutrino mass $\msne$ (the latter mainly influencing the
relative importance of $t-$channel amplitudes), we trade the usual
parameters $\mu$ and $M_2$ with : a) the lightest chargino mass
$\mcu$; b) the ratio 
\beq
r=\frac{M_2}{|\mu|} \; ,
\label{ratio}
\eeq 
 and c) $sign(\mu)$.
It will be straightforward to trace back given sets of
$(\mcu,r,sign(\mu))$ coordinates in the $(\mu,M_2)$ space of the kinematically
allowed regions in  Fig.~\ref{tre}.
\section{Cross Section Evaluation}
In this section, we present the $\hcc$ matrix element. 
As anticipated in Section 2, our analysis includes the complete  
set of 13 Feynman diagrams presented in Figs.~\ref{uno} and \ref{due}.

The matrix elements corresponding to the amplitudes $A_1,\dots,A_8$ 
in Fig.~\ref{uno} are :
\bea
{\mc M_1}&=& \f{ige^2}{k^2+i\epsilon}\bar u^\chi_{s_1}(q_1)
\left(C^L_{11}P_L+C^R_{11}P_R\right)
\f{(\not{\!q_3}+M_1)}{q_3^2-M_1^2+i\epsilon}\gamma^\mu v^\chi_{s_2}(q_2)
\bar v^e_{r_1}(p_1)
\gamma_\mu u^e_{r_2}(p_2)\nn \\
{\mc M_2}&=& \f{ige^2}{k^2+i\epsilon}\bar u^\chi_{s_1}(q_1) \gamma^\mu
\f{(-\!\!\not{\!q_4}+M_1)}{q_4^2-M_1^2+i\epsilon}
\left(C^L_{11}P_L+C^R_{11}P_R\right) v^\chi_{s_2}(q_2)
\bar v^e_{r_1}(p_1)
\gamma_\mu u^e_{r_2}(p_2)\nn \\
{\mc M_3}&=& \f{-ig^3}{4\cos^2\theta_w (k^2-M_Z^2+i\epsilon)}
\bar u^\chi_{s_1}(q_1)\left(C^L_{11}P_L+C^R_{11}P_R\right)
\f{(\not{\!q_3}+M_1)}{q_3^2-M_1^2+i\epsilon}\gamma^\mu \nn \\
&& \times\left( O^L_{11}P_L+O^R_{11} P_R \right) v^\chi_{s_2}(q_2)
\left( g_{\mu\nu}-\f{k_\mu k_\nu}{M_Z^2}\right) \bar v^e_{r_1}(p_1)
\gamma^\nu(g_V-\gamma_5) u^e_{r_2}(p_2)\nn \\
{\mc M_4}&=&\f{-ig^3}{4\cos^2\theta_w (k^2-M_Z^2+i\epsilon)}
\bar u^\chi_{s_1}(q_1)\gamma^\mu
\left( O^L_{11}P_L+O^R_{11} P_R \right) \f{(-\!\!\not{\!q_4}+M_1)}
{q_4^2-M_1^2+i\epsilon}\nn \\
&&\times\left(C^L_{11}P_L+C^R_{11}P_R\right) v^\chi_{s_2}(q_2)
\left( g_{\mu\nu}-\f{k_\mu k_\nu}{M_Z^2}\right) \bar v^e_{r_1}(p_1)
\gamma^\nu (g_V-\gamma_5)u^e_{r_2}(p_2)\nn \\
{\mc M_5}&=& \f{-ig^3}{4\cos^2\theta_w (k^2-M_Z^2+i\epsilon)}
\bar u^\chi_{s_1}(q_1)\left(C^L_{12}P_L+C^R_{12}P_R\right)
\f{(\not{\!q_3}+M_2)}{q_3^2-M_2^2+i\epsilon}\gamma^\mu \nn \\
&& \times\left( O^L_{21}P_L+O^R_{21} P_R \right) v^\chi_{s_2}(q_2)
\left( g_{\mu\nu}-\f{k_\mu k_\nu}{M_Z^2}\right) \bar v^e_{r_1}(p_1)
\gamma^\nu(g_V-\gamma_5) u^e_{r_2}(p_2)\nn \\
{\mc M_6}&=&\f{-ig^3}{4\cos^2\theta_w (k^2-M_Z^2+i\epsilon)}
\bar u^\chi_{s_1}(q_1)\gamma^\mu
\left( O^L_{12}P_L+O^R_{12} P_R \right) \f{(-\!\!\not{\!q_4}+M_2)}
{q_4^2-M_2^2+i\epsilon}\nn \\
&&\times\left(C^L_{21}P_L+C^R_{21}P_R\right) v^\chi_{s_2}(q_2)
\left( g_{\mu\nu}-\f{k_\mu k_\nu}{M_Z^2}\right) \bar v^e_{r_1}(p_1)
\gamma^\nu(g_V-\gamma_5) u^e_{r_2}(p_2)\nn \\
{\mc M_7}&=&\f{ig^3M_Z\sin{(\beta-\alpha)}}{4\cos^3\theta_w}
\bar u^\chi_{s_1}(q_1)\gamma^\mu
\left( O^L_{11}P_L+O^R_{11} P_R \right) v^\chi_{s_2}(q_2)\nn \\
&&\times \f{(g_{\mu\nu}-q_\mu q_\nu /M_Z^2)}
{(q^2-M_Z^2+i\epsilon)} \f{(g^{\nu\sigma}-k^\nu k^\sigma /M_Z^2)}
{(k^2-M_Z^2+i\epsilon)}\bar v^e_{r_1}(p_1)
\gamma_\sigma (g_V-\gamma_5) u^e_{r_2}(p_2) \nn \\
{\mc M_8}&=&\f{ig^3\cos{(\alpha-\beta)}}{8\cos^2\theta_w}
\bar u^\chi_{s_1}(q_1)
\left( {C^{A,L}_{11}}P_L+{C^{A,R}_{11}} P_R \right) v^\chi_{s_2}(q_2)\nn \\
&&\times \f{(q_\mu -h_\mu)}{(q^2-M_A^2+i\epsilon)} 
\f{(g^{\mu\nu}-k^\mu k^\nu /M_Z^2)}
{(k^2-M_Z^2+i\epsilon)}\bar v^e_{r_1}(p_1)
\gamma_\nu (g_V-\gamma_5) u^e_{r_2}(p_2) \; .
\label{matrix}
\vspace{0.8cm}
\eea
The matrix elements corresponding to the amplitudes $A_9,\dots,A_{13}$ 
in Fig.~\ref{due} are instead:
\bea
\vspace{0.4cm}
{\mc M_9}&=&\f{ig^3|V_{11}|^2}{q_5^2-M_{\tilde\nu}^2}
\bar v^e_{r_1}(p_1) P_L u^\chi_{s_1}(q_1)
\bar v^\chi_{s_2}(q_2)
\left( C_{11}^LP_L+C_{11}^R P_R \right)
\f{(-\!\!\not{\!q_4} +M_1)}{q_4^2-M_1^2+i\epsilon} 
P_R\bar u^e_{r_2}(p_2) \nn\\
\vspace{0.3cm}
{\mc M_{10}}&=&\f{ig^3|V_{11}||V_{21}|}{q_5^2-M_{\tilde\nu}^2}
\bar v^e_{r_1}(p_1) P_L u^\chi_{s_1}(q_1)
\bar v^\chi_{s_2}(q_2)
\!\left( C_{21}^LP_L+C_{21}^R P_R \right)\!
\f{(-\!\!\not{\!q_4} +M_2)}{q_4^2-M_2^2+i\epsilon} 
P_R\bar u^e_{r_2}(p_2)\!\! \nn\\
\vspace{0.3cm}
{\mc M_{11}}&=&\f{ig^3|V_{11}|^2}{q_6^2-M_{\tilde\nu}^2}
\bar v^e_{r_1}(p_1) P_L
\f{(\not{\!q_3} +M_1)}{q_3^2-M_1^2+i\epsilon} 
\left( C_{11}^LP_L+C_{11}^R P_R \right)
u^\chi_{s_1}(q_1)\bar v^\chi_{s_2}(q_2)
P_R\bar u^e_{r_2}(p_2) \nn\\
\vspace{0.3cm}
{\mc M_{12}}&=&\f{ig^3|V_{11}||V_{21}|}{q_6^2-M_{\tilde\nu}^2}
\bar v^e_{r_1}(p_1) P_L
\f{(\not{\!q_3} +M_2)}{q_3^2-M_2^2+i\epsilon} 
\!\left( C_{21}^LP_L+C_{21}^R P_R \right)\!
u^\chi_{s_1}(q_1)\bar v^\chi_{s_2}(q_2)
P_R\bar u^e_{r_2}(p_2)\!\! \nn\\
\vspace{0.3cm}
{\mc M_{13}}&=&\f{ig^3M_W\sin(\alpha+\beta)|V_{11}|^2}
{2\cos^2\theta_w(q_6^2-M_{\tilde\nu}^2)(q_5^2-M_{\tilde\nu}^2)}
\bar v^e_{r_1}(p_1) P_L
u^\chi_{s_1}(q_1)\bar v^\chi_{s_2}(q_2)
P_R\bar u^e_{r_2}(p_2) \; .
\label{matrixdue}
\vspace{0.5cm}
\eea
In Eqs.~(\ref{matrix}) and (\ref{matrixdue}), we define

$$\;\;k=p_1+p_2=q_1+q_2+h, \;\;\;q_3=q_1+h, \;\;\;q_4=q_2+h, $$ 
$$\;\;q=p_1+p_2-h,\;\;\; q_5=q_1-p_1,\;\;\; q_6=p_2-q_2~.
$$
and
$\;\;M_{1,2}=m_{\tilde\chi_{1,2}^\pm}$.

All external momenta are defined in Figs.~\ref{uno} and \ref{due}, 
as flowing from the left to the right, 
and different couplings in Eqs.~(\ref{matrix}) and (\ref{matrixdue}) 
are defined in
Appendix A. The lower indices of the spinors $u,v$ refer to the 
particle spin.

We squared, averaged over the initial spin, and summed over the final spin
the sum of the matrix elements in Eqs.~(\ref{matrix}) 
and (\ref{matrixdue}) with the help of FORM
\cite{form}.
Then, one can perform a double analytic integration over 
the phase-space variables according to the procedure described in Appendix B.
This would allow  to obtain an exact {\it analytic} expression for the 
Higgs-boson momentum distribution
\beq
  E_h\frac{d\sigma}{d^3\mathbf{h}}=
  \f{\beta}{s(4\pi)^5} \int_{-1}^1 d\cos\vartheta \; \int_0^{2\pi}d\varphi\;
  |\overline{\mathcal{M}}|^2\;= f(p_1,p_2,h)\; .
\label{dh}
\eeq
The notation is according to Appendix B, and $\mathcal{M}
=\sum^{13}_{i=1}\mathcal{M}_i\;$.
\\
In our computation,
we performed instead a completely numerical integration of the
squared matrix element in order to obtain total cross sections.
The complete code, 
including the analytic expression of the squared amplitude
and the numerical integration routine for the
evaluation of the total cross section, is available from the authors' 
e-mail addresses.

In order to check our cross section computation, we compared 
our numerical results  with the cross sections 
evaluated by CompHEP
\cite{comphep} on the basis of the same
set of Feynman diagrams, and the same input parameters.
We found complete agreement by varying the MSSM parameters in all 
the relevant range.
\section{Total Cross Sections}
\begin{figure}[th]
\vspace{-1.5cm}
\centerline{\epsfxsize=4.truein \epsfbox{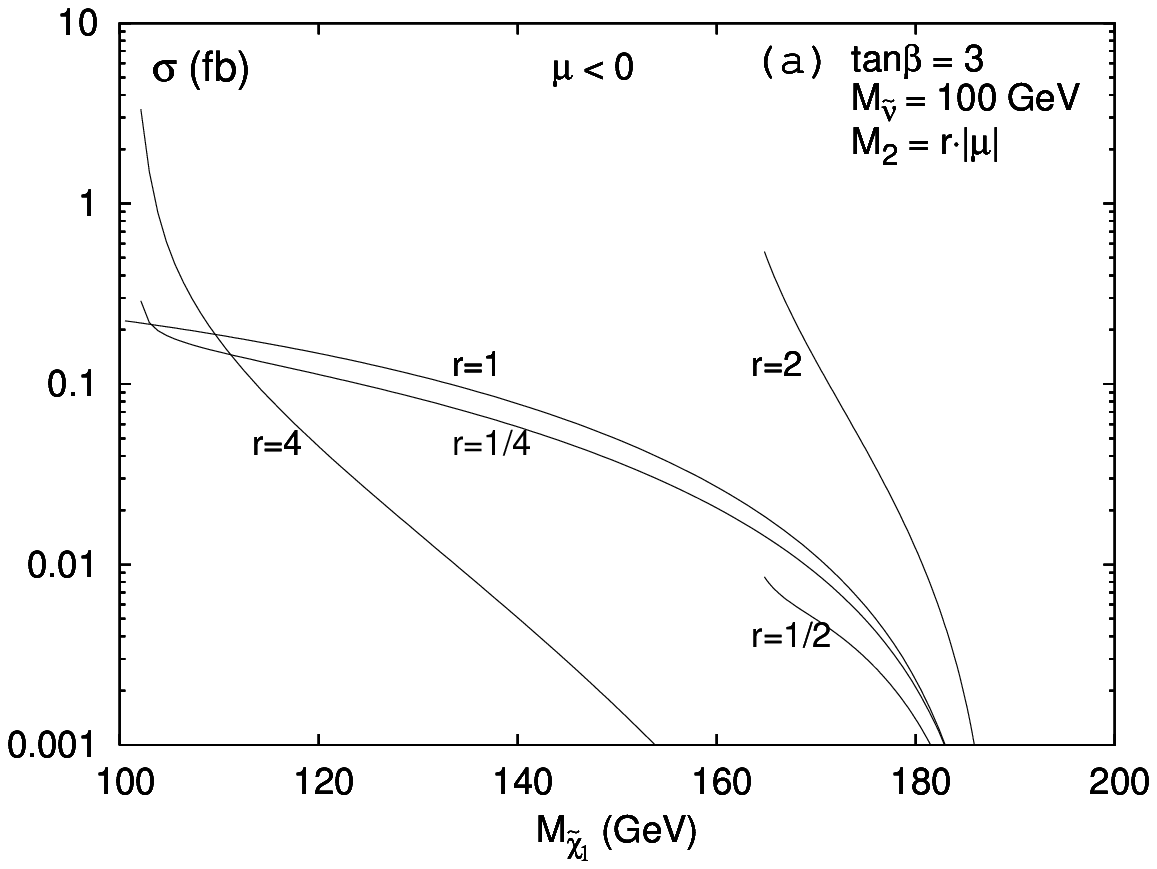}
\hspace{-1cm}\epsfxsize=4.truein \epsfbox{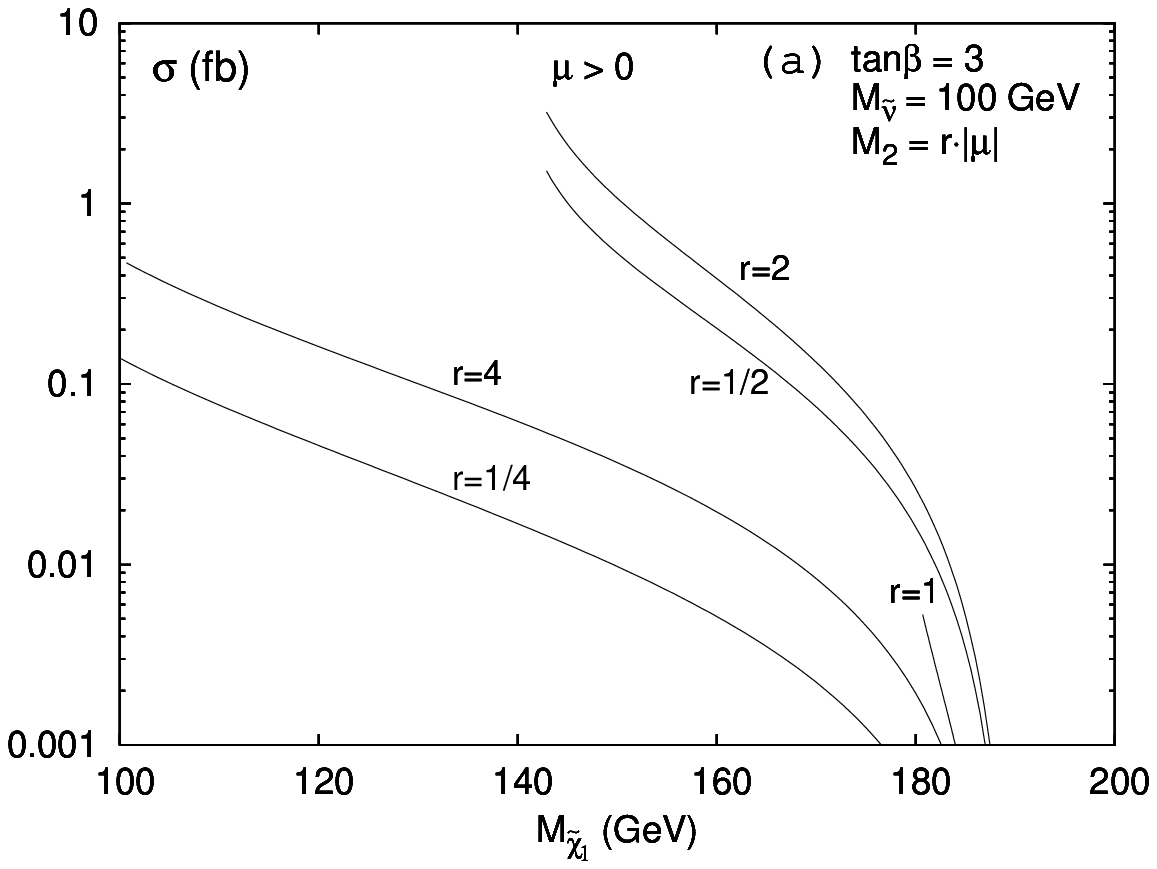}}
\vspace{0.3cm}
\centerline{\epsfxsize=4.truein \epsfbox{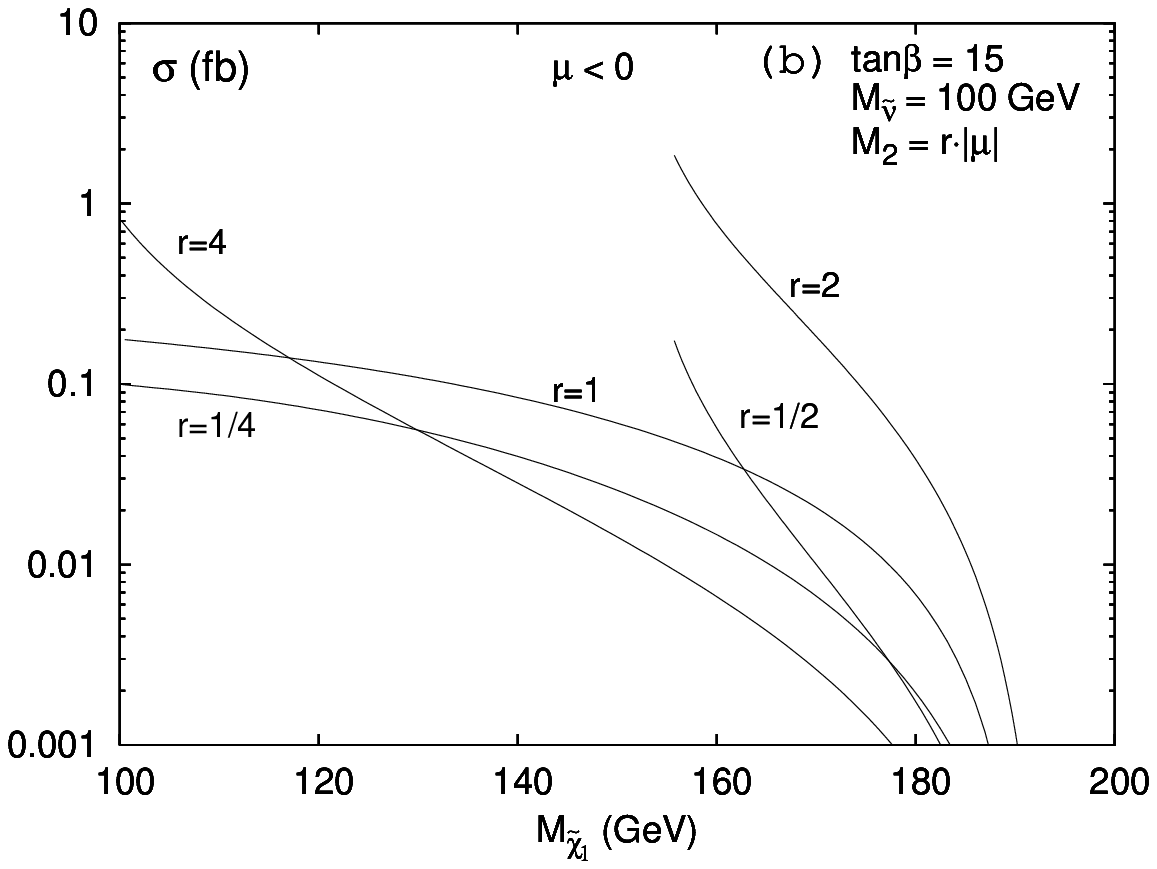}
\hspace{-1cm}\epsfxsize=4.truein \epsfbox{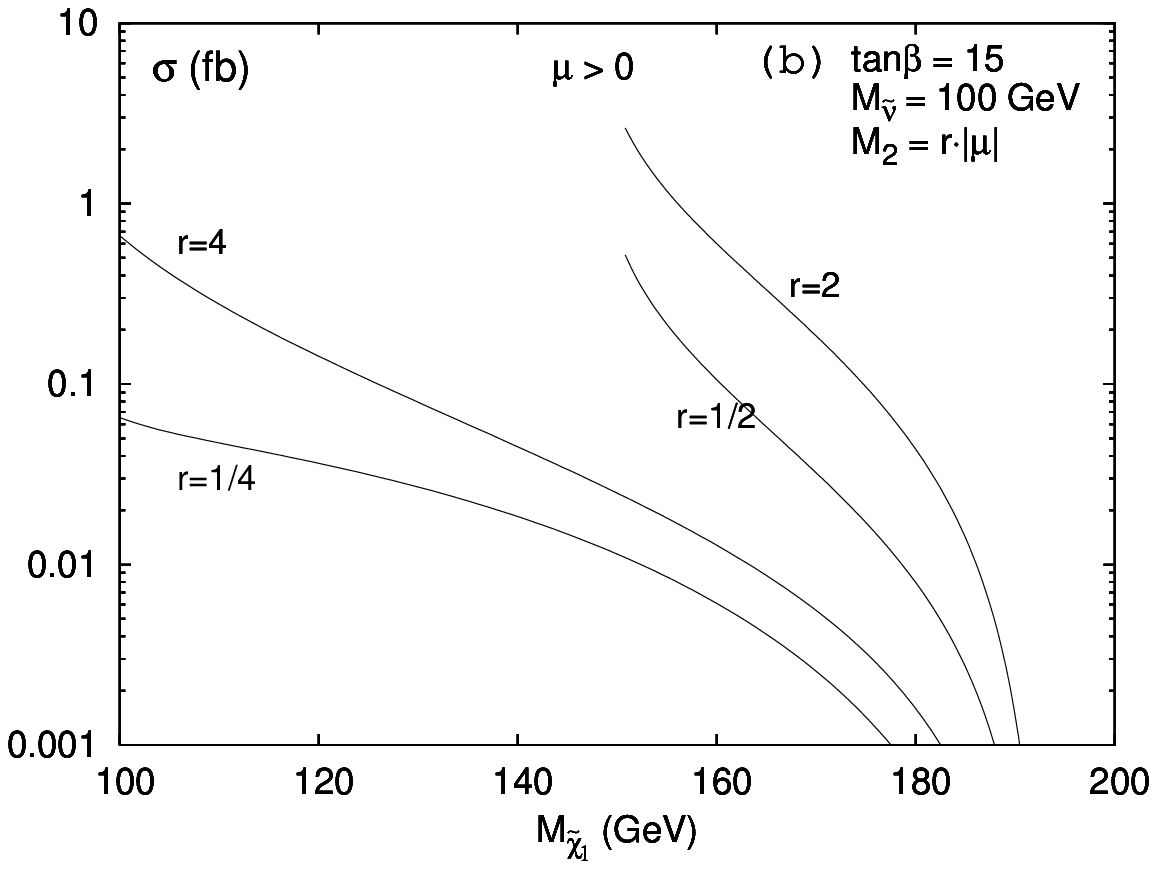}}
\vspace{0.3cm}
\centerline{\epsfxsize=4.truein \epsfbox{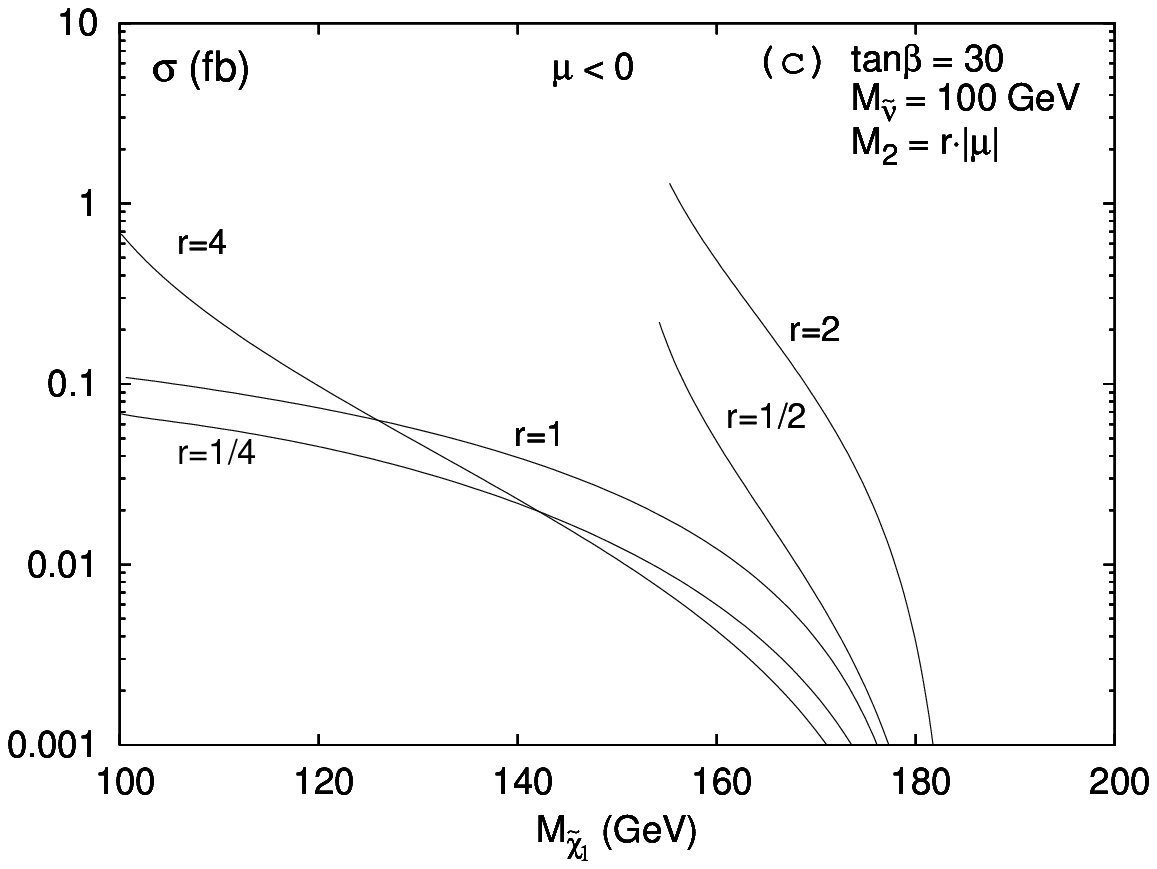}
\hspace{-1cm}\epsfxsize=4.truein \epsfbox{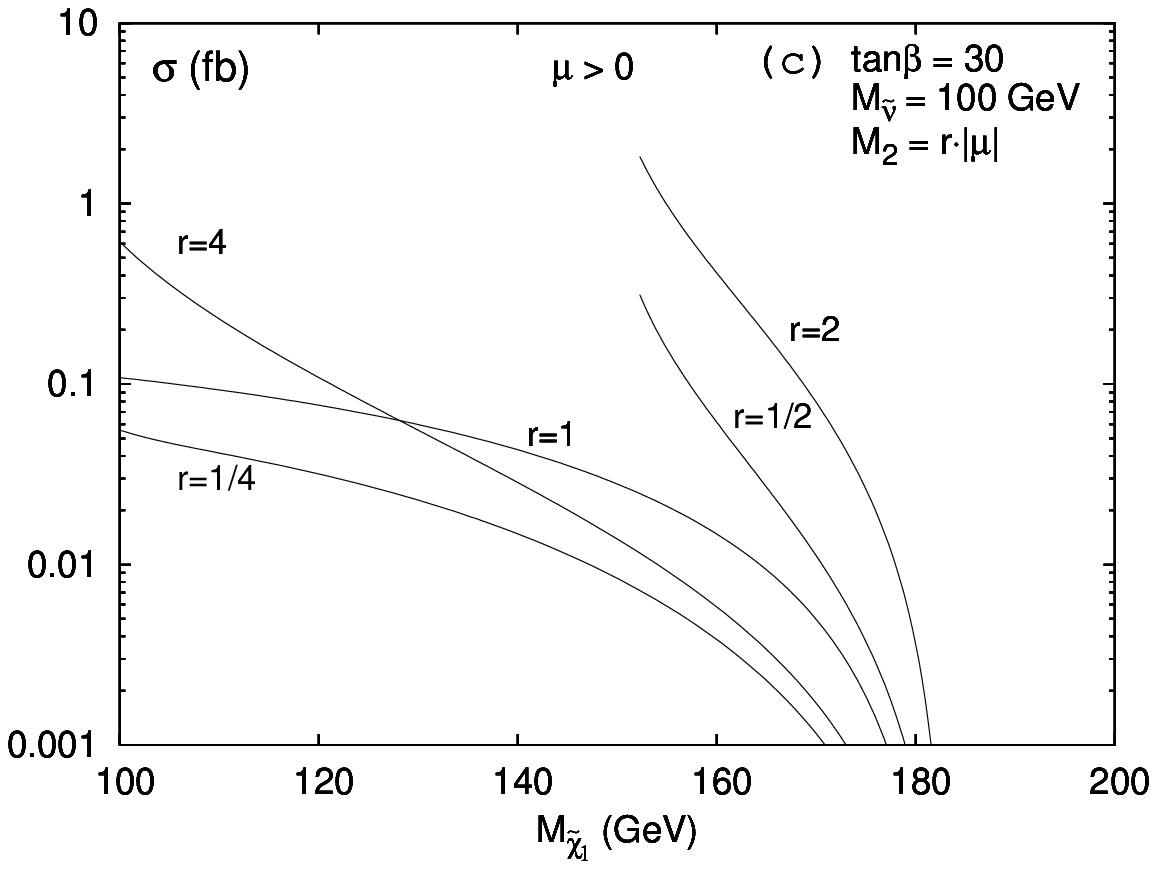}}
\vspace{-0.cm}
\caption{
\small
Total cross section for
$\hcc$ at $\sqrt s=500~GeV$, for $\tan\beta=3,15,30$, $\maa=500~GeV$, 
and $\msne=100~GeV$.
}
\label{snulight}
\end{figure}

\begin{figure}[th]
\vspace{-1.5cm}
\centerline{\epsfxsize=4.truein \epsfbox{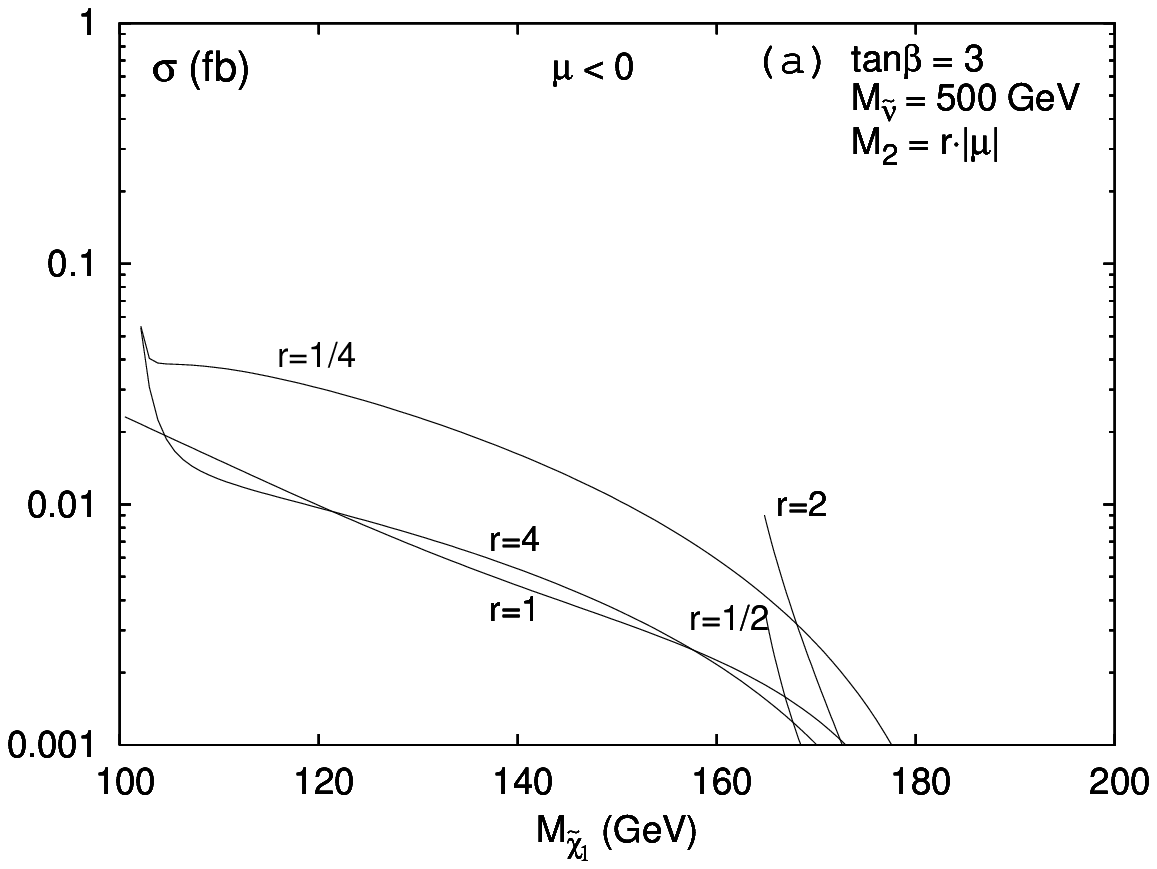}
\hspace{-1cm}
\epsfxsize=4.truein  \epsfbox{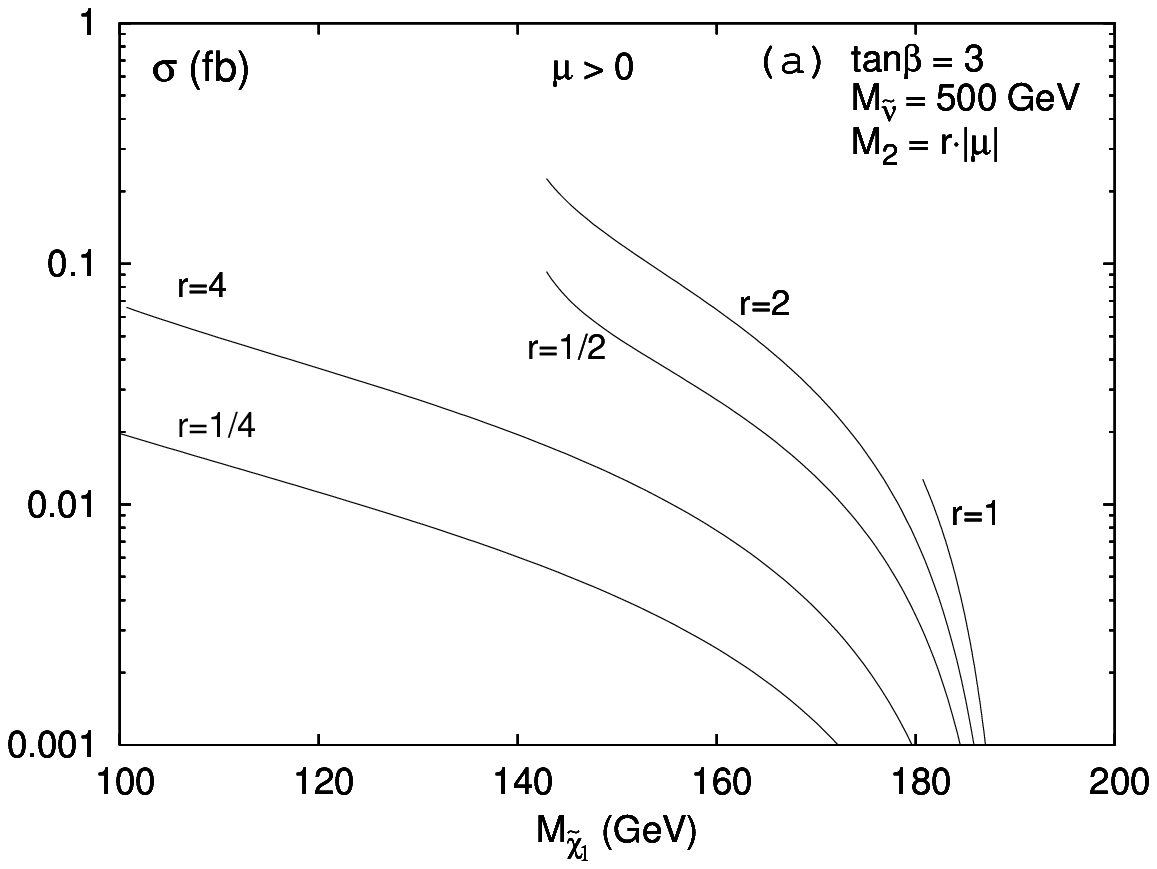}}
\vspace{0.3cm}
\centerline{\epsfxsize=4.truein \epsfbox{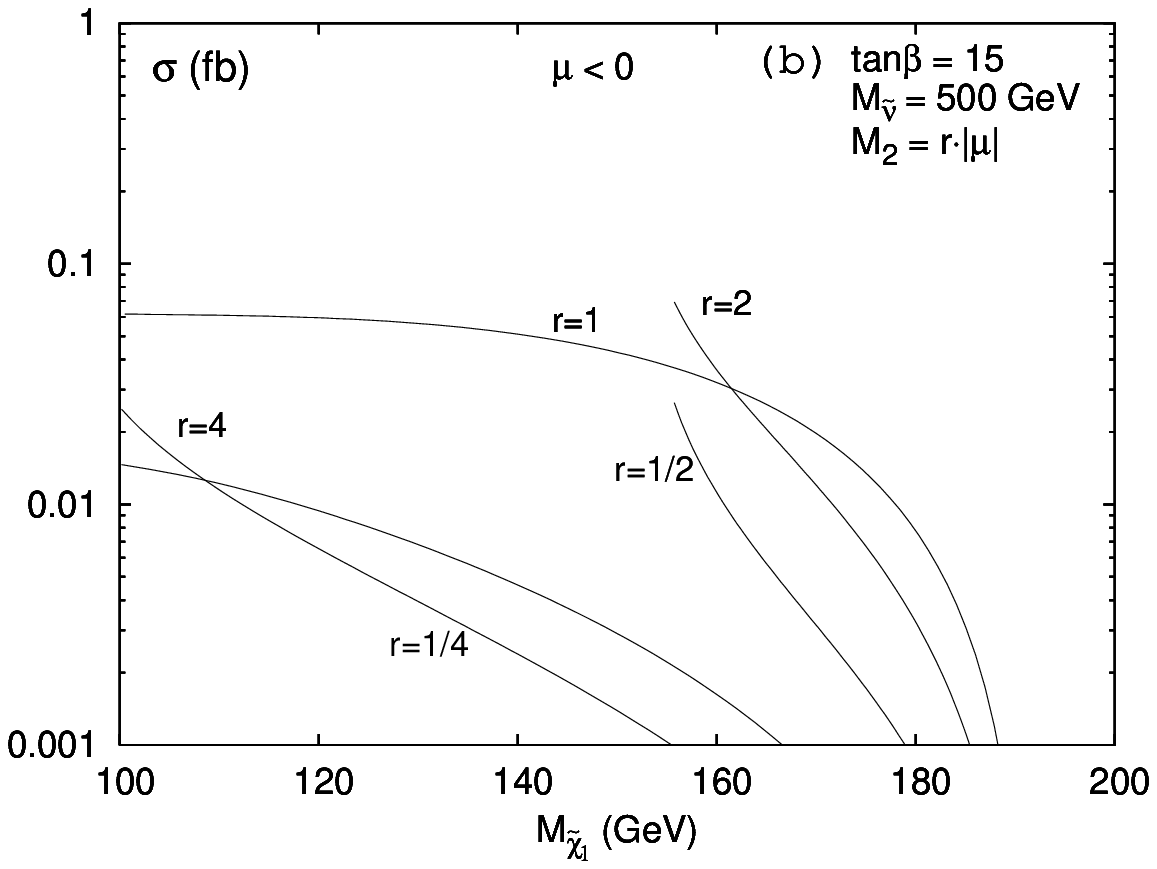}
\hspace{-1cm}\epsfxsize=4.truein \epsfbox{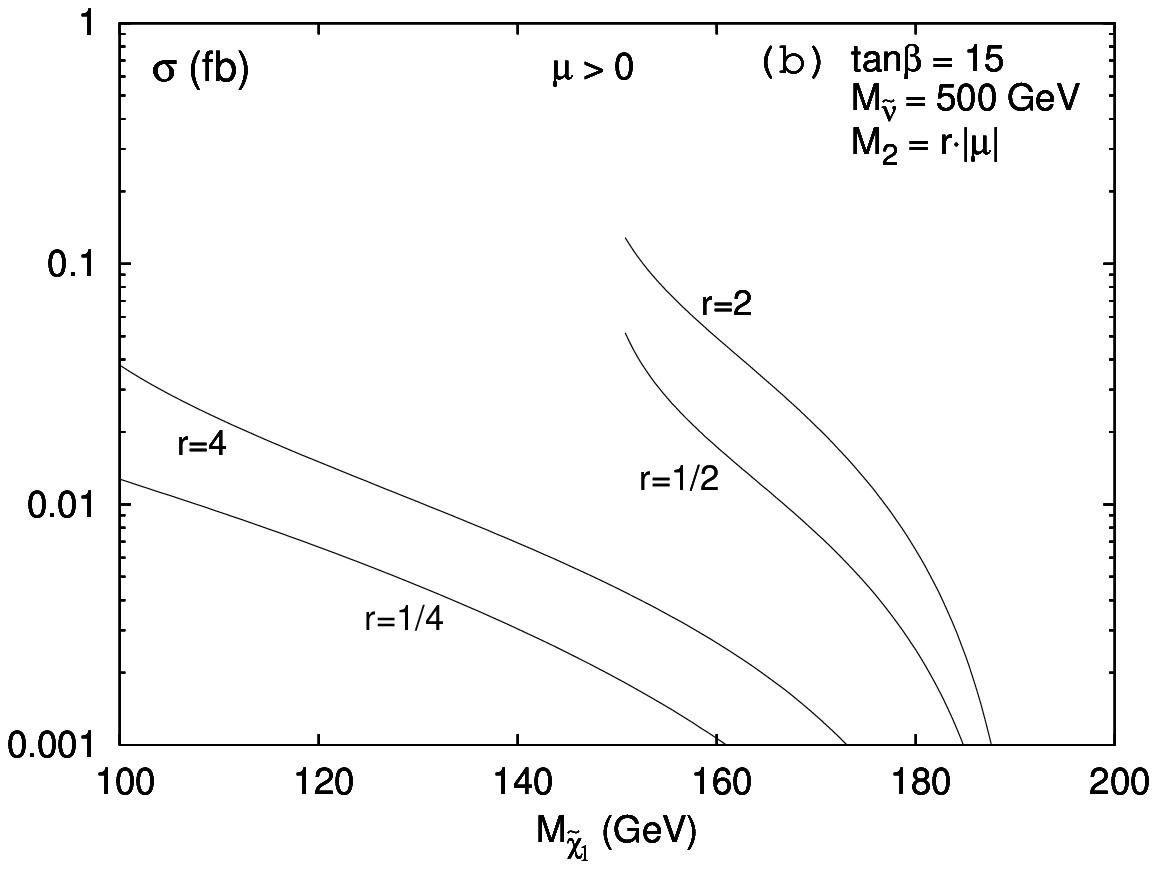}}
\vspace{0.3cm}
\centerline{\epsfxsize=4.truein \epsfbox{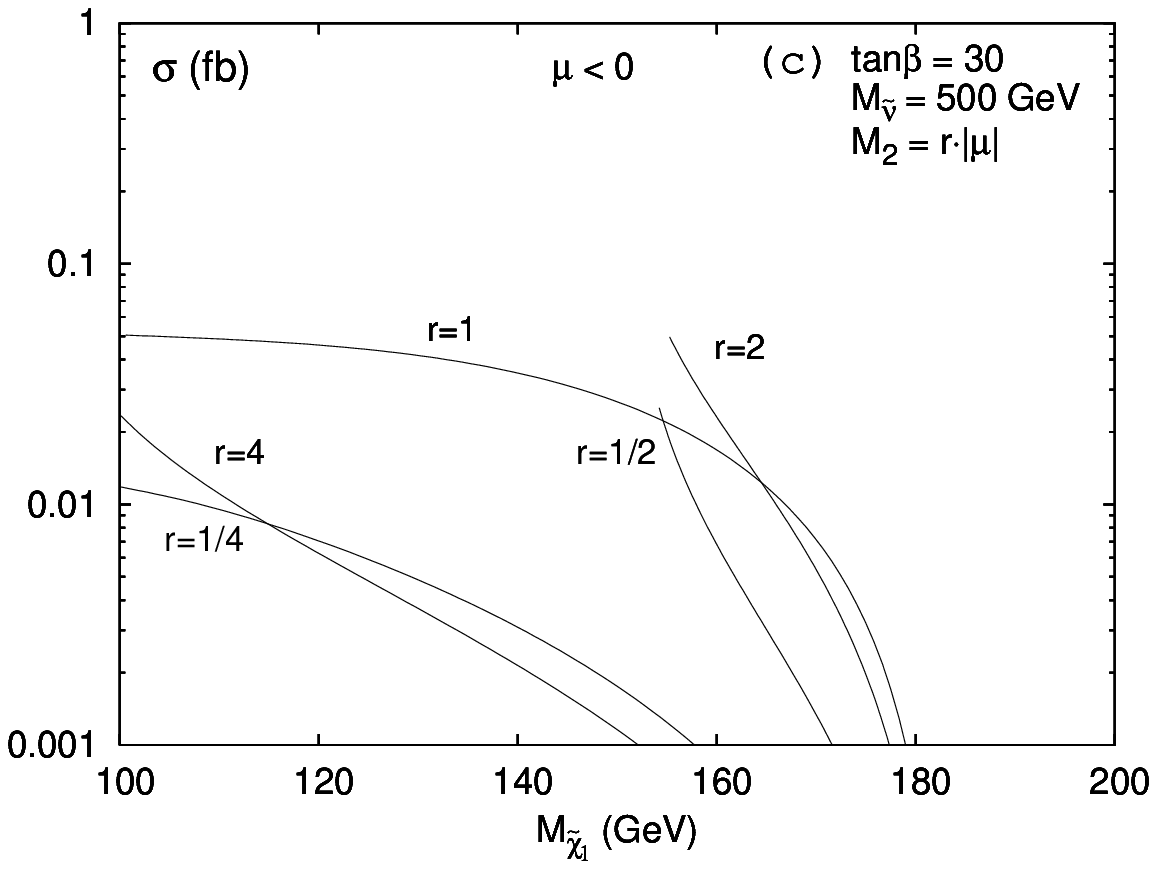}
\hspace{-1cm}\epsfxsize=4.truein \epsfbox{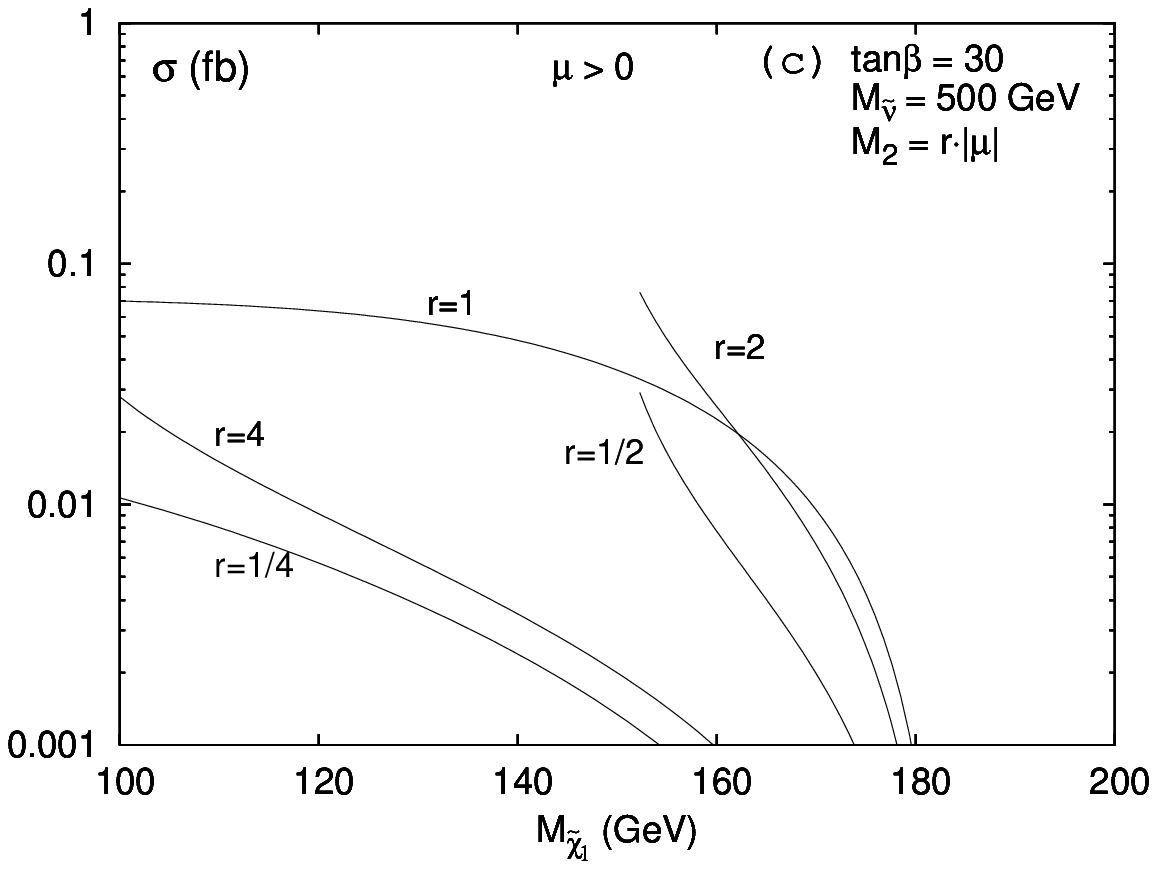}}
\vspace{-0.cm}
\caption{
\small
Total cross sections for
$\hcc$ at $\sqrt s=500~GeV$, for $\tan\beta=3,15,30$, $\maa=500~GeV$, 
and $\msne=500~GeV$.
}
\label{snuheavy}
\end{figure}

In Figs.~\ref{snulight} and \ref{snuheavy}, we show the total cross sections for
the process $\hcc$ at $\sqrt s=500~GeV$,
in the three scenarios 
{\bf a, b, c} defined in Section 2. Fig.~\ref{snulight}  assumes
 a quite light electron sneutrino ($\msne=100$ GeV), while Fig.~\ref{snuheavy}
assumes a heavier  sneutrino ($\msne=500$ GeV).
Cross sections are shown as functions of $\mcu$ at different  values
of the ratio $r=M_2/|\mu|$ (i.e., $r=1/4, 1/2, 1, 2, 4$). 
The three plots on the right (left)
part  of each figure refer to the $\mu>0$ ($\mu<0$) case.
\clearpage
In each plot, the allowed range for $\mcu$ depends  
on the value of $r$. The variation of this range versus the basic parameters
can be easily extrapolated from Fig.~\ref{tre}, keeping in mind
that only grey regions in Fig.~\ref{tre}  are kinematically allowed, and 
that a fixed $r$ value corresponds to a straight line passing through 
the $M_2=\mu=0$ point.
 To this end,
we recall that contours of fixed $\mcu$ in the parameter space of Fig.~\ref{tre}
are approximate hyperboles, spanning
the regions between the two curves 
referring to $\mcu^{min}=100$ GeV (dashed lines)
and $\mcu^{max}=(\sqrt s - \mh)/2$ (solid lines).

At $r\simeq 1$ and for low
and intermediate $\tb$, maximal $\mcu$ ranges are allowed 
only for negative $\mu$. 
At $r=1/2,2$, the allowed $\mcu$ range is always quite reduced by the condition
$\sqs<\mcu +m_{\ccm}$ 
(corresponding to the straight dot-dashed lines in Fig.~\ref{tre}),
that prevents the resonant production of a heavier
chargino.

We can see that, in general, a value $r\simeq 1$ (enhancing amplitudes 
depending on the $\hccc$ coupling)
not necessarily corresponds to larger cross sections with respect
to the case where $r$ is far from 1. This is mainly due to the competing
relevance of the amplitudes involving the $\hcccd$ coupling.
For instance, the dominance of the $r\simeq 2$ cross section on 
the $r \simeq 1$ cross section  for a light $\msne$ [cf. Fig.~\ref{snulight}], 
that is not present for a heavy $\msne$ (cf.
Fig.~\ref{snuheavy}), is due to the relative
importance of $t-$channel amplitudes involving the heavy chargino
(cf. diagrams $A_{10}$ and $A_{12}$ in Fig.~\ref{due}).
Indeed, a value $r>1$ (i.e., $M_2>\mu$) tends to  increase (decrease)
the gaugino component of the heavy (light) chargino. 

As a consequence, the sensitivity to the $\hccc$ coupling in a measurement
of the $\hcc$ total cross section will very much depend on the actual
values of the MSSM parameters, that determine the relative
importance of the amplitudes depending on the $\hccc$ vertex.

As far as the magnitude of production rates is concerned,  
for a light sneutrino 
(cf. Fig.~\ref{snulight}) it can reach a few fb's even for quite heavy
$\mcu$ ($\mcu\simeq 150$ GeV).
The typical production cross section is  
(not too close to the kinematical saturation 
of the phase-space)  of the order of 0.1 fb.
\\
For a heavy sneutrino (cf. Fig.~\ref{snuheavy}), cross sections are in general
depleted by an order of magnitude, apart from the case $r\simeq 1$ that,
at intermediate and large $\tb$, is quite insensitive to the 
$\msne$ increase.

Assuming an integrated luminosity of $1$ ab$^{-1}$ at the ILC, 
the $\hcc$ event number 
is expected to be in the range $10\div 10^3$ for a wide part of the relevant
MSSM parameter space.

In the next section, we will discuss the possibility of an experimental 
determination of the
$\hccc$ coupling through a  measurement
of the total event number
for $\hcc$ at $\sqrt s=500~GeV$.

\section{Higgs-Chargino coupling determination}
In this section, we discuss the potential of a measurement of the
total event rate for $\hcc$ at $\sqrt s=500~GeV$ for determining the
light Higgs boson coupling to charginos.

Some background for the present reaction is expected from the associated 
production of a light Higgs and electroweak vector bosons. We do not analyze 
the background in this paper. We anyhow expect that in the clean environment
of  $\eepm$ collisions the latter will be in general easily distinguishable 
on the basis
of the kinematical characteristics of the final state.
\\
In our analysis we will assume that the precision that can be achieved from
a cross section measurement will be given by the statistical error 
$\tilde \sigma$ on the cross section. 
For instance, given an integrated luminosity of  $1$ ab$^{-1}$ at the ILC, 
a cross section of 1 (0.1) fb will be affected by a statistical error
of $\tilde \sigma \simeq \,3\; (10)$ \% (corresponding to 1000 (100) events
observed).

Our strategy assumes that, before performing the present analysis,
all the basic MSSM parameter will have previously been
 measured through higher-rate
supersymmetric particle production processes (typically pair production of 
supersymmetric partners).
Our aim is to check the theoretical consistency of 
a future experimental determination of the 
coupling $\hccc$ through $\hcc$, by comparing its value with 
the MSSM predictions.

In our study, we concentrate on two different frameworks.
\\
The first assumes that the direct decay
$$
\ccp \to \cp h
$$
is allowed by phase-space (dark-grey regions
in Fig.~\ref{tre}).
Correspondingly,  a direct measurement of the $\hcccd$ 
coupling will be possible through the $\ccp \to \cp h$ decay rate.
We will also assume that the result of this measurement is consistent with
the MSSM.
Then, we will perform a one-variable analysis of the production rate,
by studying the variation of the $\hcc$ cross section versus
a possible change in the $\hccc$ coupling with respect to its MSSM value. 
We quantify the latter change through 
the parameter $\alpha_1$, as follows
\beq
  \mc L_{h^0\tilde \chi^+_1\tilde \chi^-_1} \to\;
\alpha_1\;\mc L_{h^0\tilde \chi^+_1\tilde \chi^-_1}=\;
\alpha_1\;g\;\overline{\tilde\chi}_1(x)
\;(C_{11}^{L}P_L+C_{11}^{R}P_R)\;
\tilde \chi_1(x)\;h(x)\; .
\label{alphau}
\eeq
Hence, $\alpha_1$ modifies by a total (real) normalization 
the $\hccc$ coupling in the MSSM Lagrangian (cf. Appendix A).
\\
The second framework assumes that the direct decay
$$
\ccp \to \cp h
$$
in not allowed by phase-space (light-grey regions
in Fig.~\ref{tre}).
In this case, the $\hcccd$ coupling (that also enters the $\hcc$ process)
will not be determined
through the $\ccp$ decays.
Then, we perform a two-variable analysis of the production rate,
by introducing a second parameter $\alpha_2$,
governing a possible change in the normalization
of the $\hcccd$ coupling
\beq
 \mc L_{h^0\tilde \chi^+_2\tilde \chi^-_1} \to \;
\alpha_2\;\mc L_{h^0\tilde \chi^+_2\tilde \chi^-_1}\;=\;
\alpha_2 \;g\;\overline{\tilde \chi}_1(x)
\;(C_{12}^{L}P_L+C_{12}^{R}P_R)\;
\tilde \chi_2(x)\;h(x)\; .
\label{alphad}
\eeq

\begin{figure}[th]
\vspace{-1.3cm}
\centerline{\epsfxsize=4.2truein \epsfbox{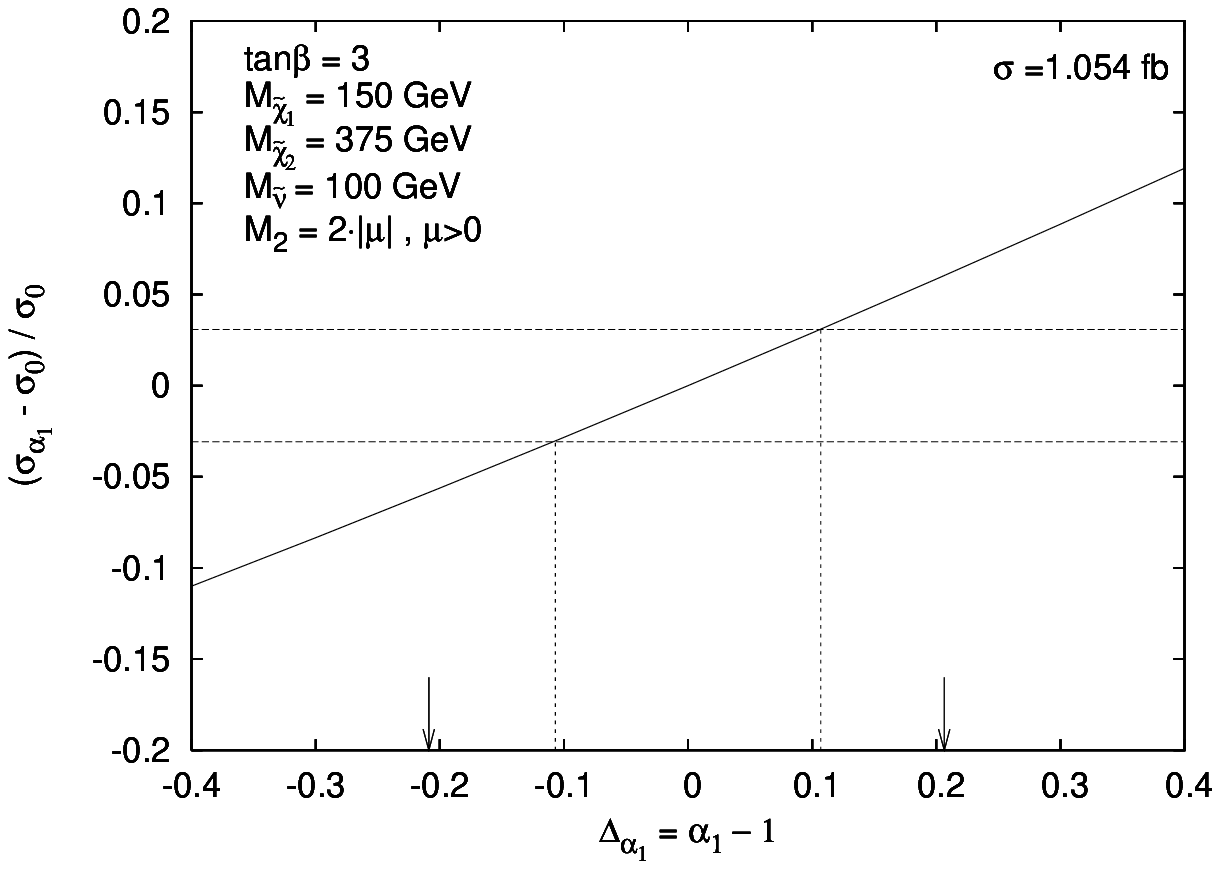}}
\centerline{\epsfxsize=4.2truein  \epsfbox{csec_a_m1150_t3_ml500+_r2.eps}}
\centerline{\epsfxsize=4.2truein \epsfbox{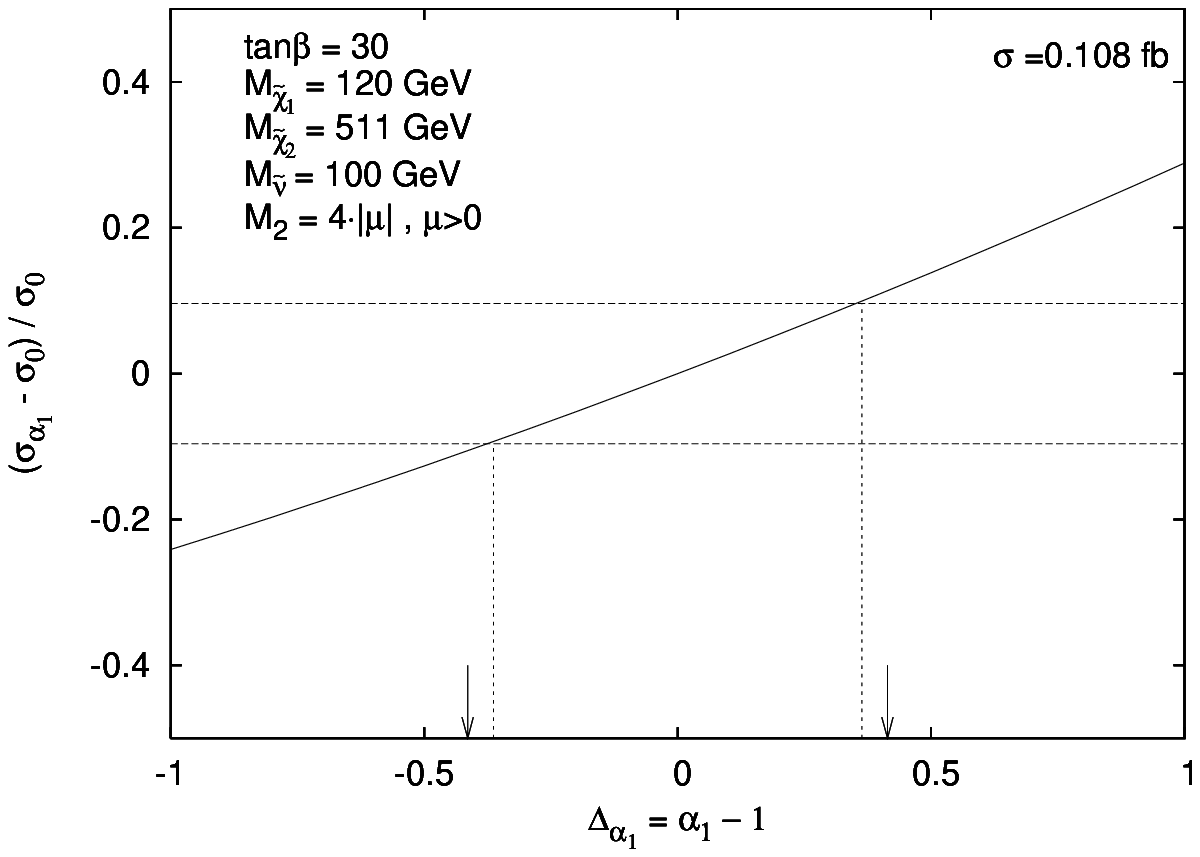}}
\vspace{0.cm}
\caption{
\small
Relative total cross section variation for 
$\hcc$ at $\sqrt s=500~GeV$, versus a change in the 
$\alpha_1$ parameter, as defined in Eq.~(\ref{alphau}), 
in three different scenarios. Arrows show
the effect of a 3 \% error  on the $\hcccd$ determination.
}
\label{sensichi}
\end{figure}

Figure~\ref{sensichi} refers to the first framework (i.e.,
allowed  $\ccp \to \cp h$ decay) in three different scenarios
corresponding to the parameters shown inside the respective plots.
The continuous lines show the relative variation 
[$(\sigma_{\alpha_1}-\sigma_{0})/\sigma_{0}$] in the total cross section
versus a change in the $\alpha_1$ parameter, as defined in Eq.~(\ref{alphau}).
The horizontal dashed lines match a variation in the cross section
corresponding to the statistical error $\pm \tilde \sigma$. Its
projection on the $\Delta_{\alpha_1}\equiv \alpha_1-1$ axis 
shows the sensitivity to a change in the $\hccc$ coupling in 
a measurement of the total rate made with an accuracy given by 
the statistical error (assuming no error on the $\hcccd$ determination
through the  $\ccp \to \cp h$ decay).
The effect of an error of 3 \% on the $\hcccd$ determination is shown by arrows
in the same plots. Of course, in scenarios where the amplitudes
containing a $\ccp$
are more relevant, this error affects more drastically the final sensitivity
to the $\hccc$ coupling.
For instance, in the second scenario in Fig.~\ref{sensichi}, 
the contribution of amplitudes
containing a $\ccp$ is negligible, and one obtains a good sensitivity
to the $\hccc$ coupling even with a moderate total cross section
($\sigma\simeq 0.12$ fb).  Indeed, in this case, $\alpha_1$ can be determined
with an error of about $\pm 7$ \%.

\begin{figure}[th]
\centerline{\epsfxsize=6.2truein \epsfbox{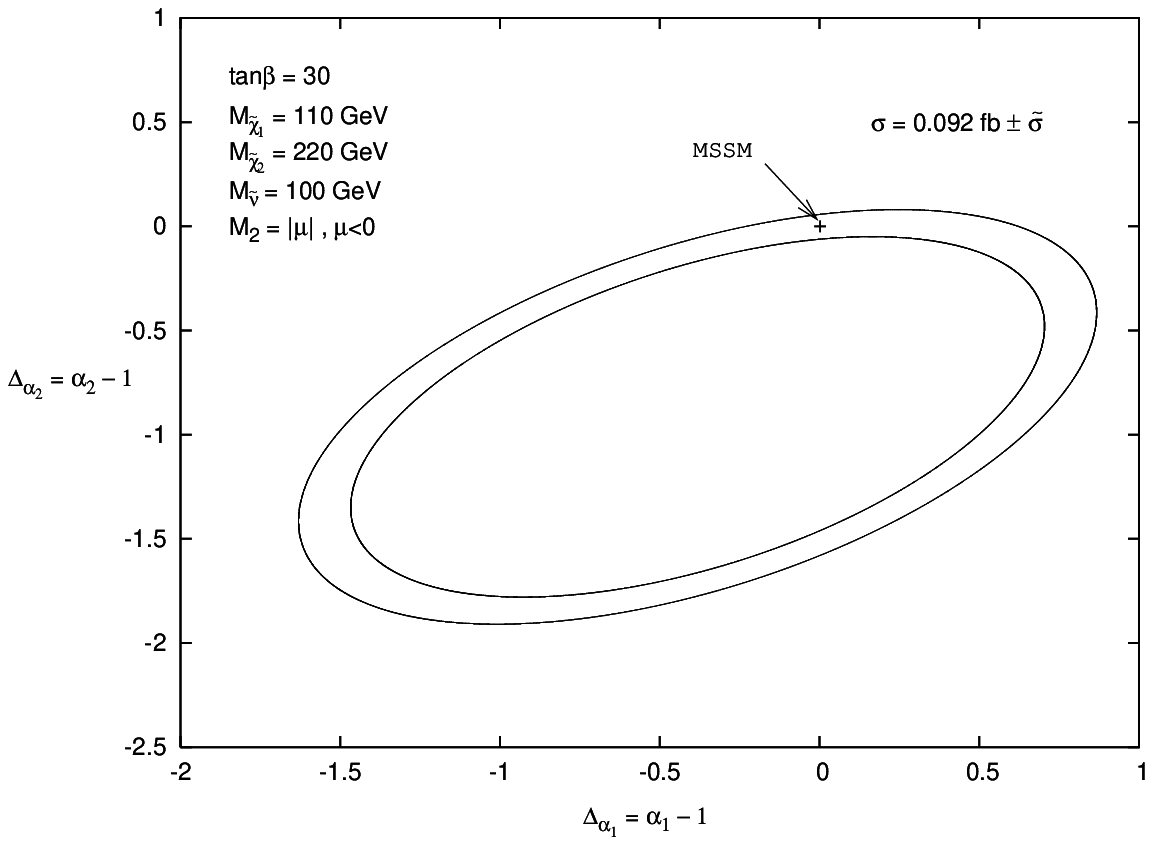}}
\caption{
\small
Total cross-section contour plot, corresponding to a variation due to the
statistical error $\tilde \sigma$, versus a change  in both the $\hccc$
coupling
 $\alpha_1$ (as defined in Eq.~(\ref{alphau})) and the $\hcccd$
coupling
 $\alpha_2$ (as defined in Eq.~(\ref{alphad})).
}
\label{sensichichi}
\end{figure}

Figure~\ref{sensichichi} refers to the more involved case where 
the $\hcccd$ coupling cannot be determined through the $\ccp \to \cp h$ decay,
that is not allowed by phase space.
In this scenario, we consider the two-dimensional dependence of the total cross
section on the variations of both the $\hccc$ and the $\hcccd$ couplings.
The area between the two cross-section contour lines corresponds
to a change around the MSSM value by the statistical error $\pm\tilde \sigma$. 
In the scenario considered in Fig.~\ref{sensichichi}, one obtains a quite 
good sensitivity to the $\hcccd$ coupling (that is better than 10 \%).
At the same time, the sensitivity to the $\hccc$ coupling is quite poor.
 
One can remark that the actual sensitivity of the  $\hcc$ 
cross section to the $\hccc$ coupling can drastically vary with the MSSM 
parameters. The real potential of the considered process for the Higgs-chargino coupling
determination will be set only after the determination of the basic MSSM
parameters, following supersymmetry discovery.
\clearpage
\newpage
\section{Conclusions}
In this paper, we analyzed the associated (non resonant)
production of a light Higgs 
boson  and a light-chargino pair  
in the MSSM, at linear colliders with $\sqrt s=500~GeV$. 

We computed the total cross section versus MSSM parameters by including the
complete set of 13 Feynman diagrams.
Cross sections up to a few fb's are found even 
 for chargino masses quite heavier than
present experimental limits. 

We discussed a possible strategy to get a first determination of the
$\hccc$ coupling through the measurement of the total rate for
$\hcc$. 
The vastly different dynamical characteristics of the various
amplitudes contributing to the $\hcc$ process make  in general
the assessment of the process potential in studying the light 
Higgs-boson coupling to charginos extremely model dependent.

We found that, in scenarios where the partial amplitudes that are directly
depending on the $\hccc$ coupling are dominant,
a determination of this coupling 
within a few percents can be reached on a purely statistical
basis, assuming an integrated luminosity of 1~ab$^{-1}$.

In case the $\ccp \to \cp h$ decay is not allowed by phase space,
a measurement of the $\hcccd$ coupling can also be obtain by
the total $\hcc$ event number, in scenarios where the partial amplitudes
depending on the $\hcccd$ coupling are relevant.

Further analysis of the measurement of the Higgs couplings to charginos, 
taking into account various systematics
and backgrounds,  will be needed in order
to assess on  more solid grounds the potential of the process
$\hcc$. 
\section*{Acknowledgments}
We thank S.~Heinemeyer, A.~Pukhov, and P.~Slavich for useful discussions.
We also thank D.~Choudury for suggesting to us  the correlation
between the Higgs coupling to charginos and the ratio $M_2/\mu$.


\section*{Appendix A: Feynman Rules}
In this Appendix we define  the couplings, parameters, 
and constants
that have been used in this paper, following the conventions in \cite{haber}.
In the evaluation of the cross section for the process $\hcc,$
we used the Feynman rules corresponding to the following
interaction Lagrangian 

\begin{itemize}
\item 
$ \;\;\mc L_{\gamma \,e^-e^+}=e\;A^\mu(x)\;\bar e(x)\;\gamma_\mu \;e(x) \; ,$

\item 
$ \;\; \mc L_{Z^0 e^-e^+}=\f g{4\cos\theta_w}\;Z^\mu(x)\;\bar e(x)\;
\gamma_\mu \;(1-4\sin^2\theta_w-\gamma_5)\;e(x) \; ,$

\item 
$ \;\; \mc L_{\gamma\,\tilde \chi^+_j\tilde \chi^-_i}=
-e\;A^\mu(x)\;\overline{\tilde \chi}_i(x)\;\gamma_\mu\; 
\tilde \chi_j(x) \;\delta_{ij}\; ,$

\item 
$ \;\; \mc L_{Z^0\tilde \chi^+_j\tilde \chi^-_i}=
\f g{\cos\theta_w}\;Z^\mu(x)\;\overline{\tilde \chi}_i(x)
\;\gamma_\mu \;(O_{ij}^{'L}P_L+O_{ij}^{'R}P_R) \;\tilde \chi_j(x)\; ,$

\item 
$ \;\;  \mc L_{h^0Z^0Z^0}=\f{g\,m_Z}{\cos\theta_w}\;Z^\mu(x)\;Z_\mu(x)\;h(x)\;
\sin(\beta-\alpha) \; ,$

\item 
$ \;\;\mc L_{h^0\tilde \chi^+_j\tilde \chi^-_i}=g\;\overline{\tilde \chi}_i(x)
\;(C_{ij}^{L}P_L+C_{ij}^{R}P_R)\;
\tilde \chi_j(x)\;h(x)\; ,$

\item 
$ \;\;\mc L_{A^0\tilde \chi^+_j\tilde \chi^-_i}=g\;\overline{\tilde \chi}_i(x)
\;(C_{ij}^{A,L}P_L+C_{ij}^{A,R}P_R)\;
\tilde \chi_j(x)\;A^0(x)\; ,$
\item 
$ \;\;\mc L_{e\tilde\nu\tilde \chi}=-g\;
\bigg\{
\overline e P_L 
\bigg(V_{11}{\tilde \chi}_1^c(x)
+V_{21}{\tilde \chi}_2^c(x)\bigg)\tilde\nu(x)+ h.c. \bigg\}\; ,$
\item 
$ \;\;\mc L_{Z^0A^0h^0}=-\f{g}{2\cos\theta_w}\;Z_\mu(x)A^0(x)(p^\mu+p^{'\mu})
\cos(\alpha-\beta)\; ,$
\item 
$ \;\;\mc L_{\tilde\nu\tilde\nu h^0}=g\f{m_w}{2\cos^2\theta_w}\;
\sin(\alpha+\beta)\tilde\nu(x)\tilde\nu(x)h(x)\; .$
\end{itemize}
where
\be P_L=\f12(1-\gamma_5),\qquad P_R=\f12(1+\gamma_5) \ee
\be O_{ij}^{'L}=-V_{i1}V_{j1}^\ast-\f12V_{i2}V_{j2}^\ast+\delta_{ij}\sin^2\theta_w 
\ee
\be O_{ij}^{'R}=-U_{i1}^\ast V_{j1}-\f12U_{i2}^\ast U_{j2}+\delta_{ij}\sin^2\theta_w 
\ee
and
\be  {\bf U}=\left(\ba{cc}
                      \cos\phi_- & \sin\phi_-\\
		     -\sin\phi_- & \cos\phi_-      \ea \right)
\ee
\be  {\bf V}=\left(\ba{cc}
                      \cos\phi_+ & \sin\phi_+\\
		     -\sin\phi_+ & \cos\phi_+      \ea \right)
\ee

\be \tan(2\phi_-)=2\sqrt 2 m_W\f{\mu \sin\beta+M_2 \cos\beta}{M_2^2-\mu^2-2m^2_W\cos(2\beta)} \ee
\be \tan(2\phi_+)=2\sqrt 2 m_W\f{\mu \cos\beta+M_2 \sin\beta}{M_2^2-
\mu^2+2m^2_W\cos(2\beta)} \;. \ee
\vskip 0.3 cm
${\bf U}$ are ${\bf V}$ are $2\times 2$ unitary matrices that diagonalize
the chargino mass matrix   ${\bf X}$
\be
{\bf U}^* {\bf X} {\bf V}^{-1} = Diag(m_{\stilde \chi_1^\pm}, m_{\stilde \chi_2^\pm})
\ee
\bea
m^2_{{\stilde \chi}_1^\pm},
m^2_{{\stilde \chi}_2^\pm}
& =
& {\f12} 
\Bigl [ (|M_2|^2 + |\mu|^2 + 2m_W^2)
\nonumber
\\
&&\mp
\sqrt{(|M_2|^2 + |\mu |^2 + 2 m_W^2 )^2 - 4 | \mu M_2 - m_W^2 \sin 2
\beta |^2 }
\Bigr ] \; .
\eea
Furthermore,
\bea C_{ij}^{L}&=&\sin\alpha \;Q_{ij}^\ast-\cos\alpha \;S_{ij}^\ast
\label{cdefil}\\
 C_{ij}^{R}&=&\sin\alpha \;Q_{ji}-\cos\alpha \;S_{ji} \label{cdefir} \\
 C_{ij}^{A,L}&=&\sin\beta \;Q_{ij}^\ast+\cos\beta \;S_{ij}^\ast \\
 C_{ij}^{A,R}&=&-\sin\beta \;Q_{ji}-\cos\beta \;S_{ji} 
\eea
\vskip 0.3 cm
where $\qquad\qquad Q_{ij}=\f1{\sqrt 2} U_{i2} V_{j1}, 
\qquad S_{ij}=\f1{\sqrt 2} U_{i1} V_{j2}\;, $\\
\vskip 0.2 cm
and 
$\qquad\qquad \tan\beta=\f{v_2}{v_1}\;,
\qquad \tan(2\alpha)=
\tan(2\beta)\left(\f{m^2_{H^0}+m^2_{h^0}}{m^2_{A^0}+m^2_Z}\right)\;$.

\section*{Appendix B : Integration of the Squared Matrix Element}
In this Appendix, we describe the details of the integration 
of the squared matrix element. In particular, we
show the procedure that can be followed in order to get
not only a completely numerical integration aimed to get 
total cross sections, but also
an analytic expression for the Higgs-boson  momentum distribution
$\dsdh$ in the process $\hcc$.
After squaring and summing/averaging over the external spins the square
of the matrix
element $\mathcal{M}=\sum^{13}_{i=1}\mathcal{M}_i\;$
obtained from Eqs.~(\ref{matrix}) and (\ref{matrixdue}) 
(we did that with the help of FORM \cite{form}),
one can perform two analytic integrations of $|\overline{\mathcal{M}}|^2$
(the squared modulus of $\mathcal{M}$ avaraged over the initial particles spin)
over the phase-space variables
in the following way.
{
\begin{figure}[ht]
  \begin{center}
\unitlength=1.0 pt
\SetScale{1.0}
\SetWidth{0.7}      
    \begin{picture}(170,150)(0,0)
      \Line(0,0)(70,30)
      \Line(70,30)(170,30)
      \Line(70,100)(70,120)
      \DashLine(15,50)(15,6){3.5}
      \DashLine(110,70)(110,6){3.5}
      \DashLine(110,6)(70,30){3.5}
      \LongArrowArc(70,30)(8,205,327)
      \LongArrowArc(70,30)(8,90,157)
      \LongArrowArcn(70,30)(8,90,43)
      \SetWidth{1}
      \LongArrow(70,30)(70,100)
      \LongArrow(70,30)(110,70)
      \LongArrow(70,30)(15,50)
      \Text(75,100)[lb]{$\mathbf{p_1}$}
      \Text(115,70)[lb]{$\mathbf{q_1}$}
      \Text(10,50)[rb]{$\mathbf{p_2}$}
      \Text(70,20)[t]{$\varphi$}
      \Text(60,45)[]{$\chi$} 
      \Text(77,47)[]{$\vartheta$}
      \Text(72,120)[lb]{\itshape z}
      \Text(170,28)[tl]{\itshape y}
      \Text(0,-1)[t,l]{\itshape x}
    \end{picture}
  \end{center}
  \caption{Angular-variables definition in the chargino-pair c.m. frame.}
  \label{rif}
\end{figure}
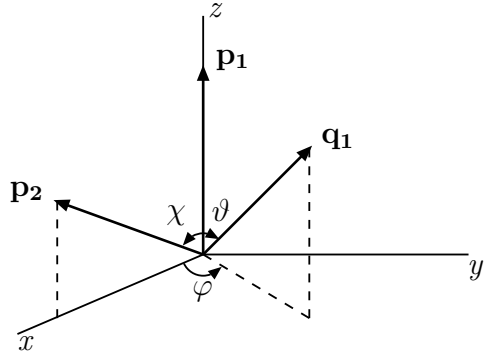
}

\noindent Starting from the momenta definition
\begin{equation}
  e^+(p_1)+e^-(p_2) \longrightarrow 
  \tilde{\chi}_1^+(q_1)+\tilde{\chi}_1^-(q_2)+
  h^0(h),
\end{equation}
and 
\begin{equation}
  \;\;p_1=(E_1,\mathbf{p_1}),\;\;\;p_2=(E_2,\mathbf{p_2}),
  \;\;\;q_1=(E_1',\mathbf{q_1}),\;\;\;q_2=(E_2',\mathbf{q_2}),
  \;\;\;h=(E_h,\mathbf{h}) \; ,
\end{equation}
the Higgs momentum distribution can be expressed as
\begin{equation}
  E_h\frac{d\sigma}{d^3\mathbf{h}}=\f1{(2\pi)^5}
\int\frac{|\overline{\mathcal{M}}|^2}
  {16s}\;\delta^4(p_1+p_2-q_1-q_2-h)\;\frac{d^3\mathbf{q_1}}{E_1'}
  \frac{d^3\mathbf{q_2}}{E_2'} \; ,
  \label{dsdheq}
\end{equation}
where $s=(p_1+p_2)^2=2(p_1p_2)$.

In order to perform analytically
the two non trivial integrations in Eq.~(\ref{dsdheq}),
 one can first express $|\overline{\mathcal{M}}|^2$ as a function
of the following five independent products of momenta
\begin{equation}
  s,\;\; (p_1h),\;\; (p_2h),\;\; (p_1q_1),\;\; (p_2q_1)\;.
\end{equation}
Then, one can  express $(p_1q_1)$ and $(p_2q_1)$ in the chargino-pair 
c.m. system 
(where $\mathbf{q_1}+\mathbf{q_2}=\mathbf{0}$) as a function of the angular
variables defined in Fig.~\ref{rif}, as follows
\begin{eqnarray}
  (p_1q_1)&=&\frac{s_1}{4}(1-\beta \cos\vartheta), \\
  (p_2q_1)&=&\frac{s_2}{4}(1-\beta \cos\vartheta \cos\chi 
  -\beta \sin\vartheta \sin\chi \cos\varphi). 
\end{eqnarray}
where
\begin{eqnarray}
  \beta&=&\sqrt{1-\frac{4M_1^2}{s+m_h^2-2(p_1h)-2(p_2h)}},\\
    \cos\chi&=&1-\frac{2s(s+m_h^2-2(p_1h)-2(p_2h))}{(s-2(p_1h))(s-2(p_2h))},  
\end{eqnarray}
and 
\beq 
s_{1,2}=s-2(p_{1,2}h).
\eeq
Then, one can write the differential cross section as
\beq
  E_h\frac{d\sigma}{d^3\mathbf{h}}=
  \f{\beta}{s(4\pi)^5} \int_{-1}^1 d\cos\vartheta \; \int_0^{2\pi}d\varphi\;
  |\overline{\mathcal{M}}|^2\; ,
\label{dsigmadh}
\eeq
and perform analytically the two angular integrations.
The result (that is a quite lengthy expression) is
a relativistic invariant function of $(p_1h)$, $(p_2h)$ and $s$.

The total cross section can be finally worked out by integrating
numerically the result of Eq.~(\ref{dsigmadh}) over the Higgs-boson 
momentum in the $\eepm$ c.m. system (where $\mathbf{p_1}+\mathbf{p_2}=0$),
\begin{equation}
  \sigma=
  2\pi\int_{{E_h}^{min}}^{{E_h}^{max}}dE_h \; \; |\mathbf h|
  \int_{-1}^{1}d\cos\theta 
 \left[E_h\frac{d\sigma}{d^3\mathbf h}\Bigl ( (p_1h),(p_2h) \Bigr )\right].
\label{intn}
\end{equation}
In Eq.(\ref{intn}), 
\be
(p_1h)=\frac{\sqrt s}2(E_h-|\mathbf h| \cos\theta), ~~~~~~~
(p_2h)=\frac{\sqrt s}2(E_h+|\mathbf h| \cos\theta),
\nonumber
\ee
with $|\mathbf h|=\sqrt{E_h^2-m_h^2}\;$ , 
$\;\;\;{E_{h}}^{min}=m_h\;,\;\;$ and 
$\;\;{E_h}^{max}=(s+m_h^2-4M_1^2)/(2\sqrt{s})\;$ .



\begin{thebibliography}{20}


\bibitem{Accomando:1997wt}
E.~Accomando {\it et al.}  [ECFA/DESY LC Physics Working Group
                  Collaboration],
Phys.\ Rept.\  {\bf 299} (1998) 1.
[arXiv:hep-ph/9705442]; \\
J.~A.~Aguilar-Saavedra {\it et al.}  [ECFA/DESY LC Physics Working Group
                  Collaboration],
arXiv:hep-ph/0106315; \\
K.~Abe {\it et al.}  [ACFA Linear Collider Working Group Collaboration],
arXiv:hep-ph/0109166; \\
T.~Abe {\it et al.}  [American Linear Collider Working Group Collaboration],
in {\it Proc. of the APS/DPF/DPB Summer Study on the Future of Particle 
Physics (Snowmass 2001) } ed. N.~Graf,
arXiv:hep-ex/0106055 ;
arXiv:hep-ex/0106056 ;
arXiv:hep-ex/0106057 ; 
arXiv:hep-ex/0106058.


\bibitem{Gunion:1989we}
J.~F.~Gunion, H.~E.~Haber, G.~L.~Kane and S.~Dawson,
{\it The Higgs Hunter's Guide},
Addison-Wesley, 1990.

\bibitem{Nilles:1983ge}
H.~P.~Nilles,
Phys.\ Rept.\  {\bf 110} (1984) 1 .
\bibitem{haber}
H.~E.~Haber and G.~L.~Kane,
Phys.\ Rept.\  {\bf 117} (1985) 75 .
\bibitem{barbieri}
R.~Barbieri,
Riv.\ Nuovo Cim.\  {\bf 11} (1988) 1.



\bibitem{Gaemers:1978jr}
K.~J.~Gaemers and G.~J.~Gounaris,
Phys.\ Lett.\ B {\bf 77} (1978) 379; \\
A.~Djouadi, J.~Kalinowski and P.~M.~Zerwas,
Mod.\ Phys.\ Lett.\ A {\bf 7} (1992) 1765 and
Z.\ Phys.\ C {\bf 54} (1992) 255; \\
J.~F.~Gunion, B.~Grzadkowski and X.~G.~He,
Phys.\ Rev.\ Lett.\  {\bf 77} (1996) 5172,
[arXiv:hep-ph/9605326] ; \\
H.~Baer, S.~Dawson and L.~Reina,
Phys.\ Rev.\ D {\bf 61} (2000) 013002,
[arXiv:hep-ph/9906419] ; \\
A.~Juste and G.~Merino,
arXiv:hep-ph/9910301 ; \\
S.~Moretti,
Phys.\ Lett.\ B {\bf 452} (1999) 338,
[arXiv:hep-ph/9902214] ; \\
A.~Denner, S.~Dittmaier, M.~Roth and M.~M.~Weber,
arXiv:hep-ph/0309274 and references therein.


\bibitem{Belanger:1998rq}
G.~Belanger, F.~Boudjema, T.~Kon and V.~Lafage,
Eur.\ Phys.\ J.\ C {\bf 9} (1999) 511,
[arXiv:hep-ph/9811334] ; \\
A.~Djouadi, J.~L.~Kneur and G.~Moultaka,
Nucl.\ Phys.\ B {\bf 569} (2000) 53,
[arXiv:hep-ph/9903218].


\bibitem{Djouadi:1997xx}
A.~Djouadi, J.~L.~Kneur and G.~Moultaka,
Phys.\ Rev.\ Lett.\  {\bf 80} (1998) 1830,
[arXiv:hep-ph/9711244].
A.~Dedes and S.~Moretti,
Phys.\ Rev.\ D {\bf 60} (1999) 015007 
[arXiv:hep-ph/9812328] and Eur.\ Phys.\ J.\ C {\bf 10} (1999) 515.
[arXiv:hep-ph/9904491] ; \\
G.~Belanger, F.~Boudjema and K.~Sridhar,
Nucl.\ Phys.\ B {\bf 568} (2000) 3,
[arXiv:hep-ph/9904348].

\bibitem{Datta:2001sh}
A.~Datta, A.~Djouadi and J.~L.~Kneur,
Phys.\ Lett.\ B {\bf 509} (2001) 299
[arXiv:hep-ph/0101353].

\bibitem{Djouadi:1996pj}
A.~Djouadi, J.~Kalinowski, P.~Ohmann and P.~M.~Zerwas,
Z.\ Phys.\ C {\bf 74} (1997) 93
[arXiv:hep-ph/9605339]; \\
A.~Bartl, H.~Eberl, K.~Hidaka, T.~Kon, W.~Majerotto and Y.~Yamada,
Phys.\ Lett.\ B {\bf 389} (1996) 538
[arXiv:hep-ph/9607388].

\bibitem{Fraas:2003cx}
H.~Fraas, F.~Franke, G.~Moortgat-Pick, F.~von der Pahlen and A.~Wagner,
arXiv:hep-ph/0303044.

\bibitem{Brignole:2001jy}
A.~Brignole, G.~Degrassi, P.~Slavich and F.~Zwirner,
Nucl.\ Phys.\ B {\bf 631} (2002) 195
[arXiv:hep-ph/0112177] and
Nucl.\ Phys.\ B {\bf 643} (2002) 79
[arXiv:hep-ph/0206101].


\bibitem{LEPSUSYWG}
LEPSUSYWG, ALEPH, DELPHI, L3 and OPAL experiments, note LEPSUSYWG/01-03.1
and note LEPSUSYWG/02-04.1
(http://lepsusy.web.cern.ch/lepsusy/Welcome.html). 


\bibitem{Baillargeon:1993iw}
M.~Baillargeon, F.~Boudjema, F.~Cuypers, E.~Gabrielli and B.~Mele,
Nucl.\ Phys.\ B {\bf 424} (1994) 343
[arXiv:hep-ph/9307225].


\bibitem{lesh}
B.~C.~Allanach {\it et al.}  [Beyond the Standard Model Working Group
                  Collaboration],
 ``Les Houches 'Physics at TeV Colliders 2003' Beyond the Standard Model
Working Group: Summary report'',
arXiv:hep-ph/0402295.
\bibitem{lephiggsMSSM} 
ALEPH, DELPHI, L3, OPAL Collaborations and the LEP Higgs Working Group,
LHWG Note 2001-4 {\it [ALEPH 2001-057, DELPHI 2001-114, L3 Note 2007, OPAL 
Technical Note TN699]}, CERN preprint 2001; \\
OPAL Collaboration, OPAL PN524, CERN preprint 2003.

\bibitem{lephiggsSM}
R.~Barate {\it et al.} 
[ALEPH, DELPHI, L3, OPAL
Collaborations and LEP Working Group for Higgs boson searches],
Phys.\ Lett.\ B {\bf 565} (2003) 61
[arXiv:hep-ex/0306033].

\bibitem{FeynHiggsFast}
S.~Heinemeyer, W.~Hollik and G.~Weiglein,
 ``FeynHiggsFast: A program for a fast calculation of masses and 
  mixing  angles in the Higgs sector of the MSSM'',
arXiv:hep-ph/0002213.

\bibitem{form}
J.A.M.~Vermaseren, {\it Symbolic Manipulation with FORM},
published by CAN (Computer Algebra Nederland), Kruislaan 413, 1098
SJ Amsterdam, 1991, ISBN 90-74116-01-9.

\bibitem{comphep}
A.~Pukhov {\it et al.},
``CompHEP: A package for evaluation of Feynman diagrams and integration  over
multi-particle phase space. User's manual for version 33'',
arXiv:hep-ph/9908288; \\
A.~Semenov,
``CompHEP/SUSY package'',
Nucl.\ Instrum.\ Meth.\ A {\bf 502} (2003) 558
[arXiv:hep-ph/0205020].

\end{thebibliography}
\end{document}